\documentclass[journal=jacsat,manuscript=article]{achemso}

\usepackage{chemformula} 
\usepackage[T1]{fontenc} 
\usepackage{multirow}
\usepackage{siunitx}

\newcommand{\Ert}{$\mathrm{Er}^{3+}$}
\newcommand{\Ybt}{$\mathrm{Yb}^{3+}$}
\newcommand{\dHo}{$^2\mathrm{H}_{11/2}$}
\newcommand{\qIq}{$^4\mathrm{I}_{15/2}$}
\newcommand{\qIo}{$^4\mathrm{I}_{11/2}$}
\newcommand{\qIt}{$^4\mathrm{I}_{13/2}$}
\newcommand{\qFn}{$^4\mathrm{F}_{9/2}$}
\newcommand{\qSt}{$^4\mathrm{S}_{3/2}$}
\newcommand{\dFs}{$^2\mathrm{F}_{7/2}$}
\newcommand{\dFc}{$^2\mathrm{F}_{5/2}$}

\newcommand{\YO}{$\mathrm{Y}_2\mathrm{O}_3$}

\author{Pauline Perrin}
\affiliation{Chimie ParisTech, PSL University, CNRS, Institut de Recherche de Chimie Paris, 75005 Paris, France }
\email{pauline.perrin@chimieparistech.psl.eu}
\author{Luiz Fernando Dos Santos}
\affiliation{Laboratorio de Materiais Luminescentes Micro e Nanoestruturados – Mater Lumen, Departamento de Química, Faculdade de Filosofia, Ciencias e Letras de Ribeir\~ao Preto, Universidade de S\~ao Paulo, Ribeir\~ao Preto, SP, Brazil}
\alsoaffiliation{Chimie ParisTech, PSL University, CNRS, Institut de Recherche de Chimie Paris, 75005 Paris, France }
\author{Diana Serrano}
\author{Alexey Tiranov}
\affiliation{Chimie ParisTech, PSL University, CNRS, Institut de Recherche de Chimie Paris, 75005 Paris, France }
\author{Jocelyn Achard}
\affiliation{LSPM, CNRS, Universite Sorbonne Paris Nord, 99 avenue JB Clement, 93460 Villetaneuse, France}
\author{Alexandre Tallaire}
\affiliation{Chimie ParisTech, PSL University, CNRS, Institut de Recherche de Chimie Paris, 75005 Paris, France }
\author{Rogéria R. Gonçalves}
\affiliation{Laboratorio de Materiais Luminescentes Micro e Nanoestruturados – Mater Lumen, Departamento de Química, Faculdade de Filosofia, Ciencias e Letras de Ribeir\~ao Preto, Universidade de S\~ao Paulo, Ribeir\~ao Preto, SP, Brazil}
\author{Philippe Goldner}
\affiliation{Chimie ParisTech, PSL University, CNRS, Institut de Recherche de Chimie Paris, 75005 Paris, France }
\email{philippe.goldner@chimieparistech.psl.eu}

\title[An \textsf{achemso} demo]
  {Interactions in Rare Earth Doped Nanoparticles: A Multi-Transition, Concentration, and Excitation Path  Analysis}

\abbreviations{IR,NMR,UV}
\keywords{American Chemical Society, \LaTeX}

\begin{document}



\begin{abstract}
Understanding and modeling energy transfer mechanisms in rare-earth-doped nanomaterials is essential for advancing luminescent technologies used in bioimaging, optical thermometry, and solid-state lasers. In this work, we investigate the photoluminescence dynamics of Yb$^{3+}$ and Er$^{3+}$ ions in Y$_2$O$_3$ nanoparticles over a wide concentration range (0.5–17\%), using both direct and upconversion excitation. Luminescence decays of green, red, and near-infrared transitions were measured and analyzed using a rate-equation model incorporating radiative and non-radiative processes, energy transfer mechanisms, and defect-related quenching. The model successfully reproduces experimental trends across most concentrations and excitation paths. This work provides a reliable and predictive framework for modeling energy transfer in rare-earth doped materials and offers valuable insights for optimizing photoluminescent properties in nanostructured systems.
\end{abstract}


\section{Introduction}

Energy transfer phenomena in rare-earth doped materials have been investigated for a long time and remain a major topic in solid-state luminescence on both theoretical aspects and applications \cite{Weber1971,huberDynamicsIncoherentTransfer1981,mangnusFiniteSizeEffectsEnergy2022}. Among many excitation/emission schemes, those enabling emission following excitation at longer wavelengths are of particular interest. They rely on energy transfers between excited states with long enough lifetimes. As the number of successive energy transfers increases, so does the energy of the populated levels. The so-called energy transfer up-conversion (ETU) \cite{auzel1966,wrightUpconversionExcitedState1976,gaiRecentProgressRare2014,duNanocompositesBasedLanthanidedoped2022} and closely related phenomena such as avalanche up-conversion \cite{leeGiantNonlinearOptical2021}, finds applications in many fields such as lasers \cite{scheifeAdvancesUpconversionLasers2004}, thermometry \cite{caixetaHighQuantumYieldUpconvertingEr2020}, bioimaging \cite{zengVisualizationIntraneuronalMotor2019}, or super-resolution fluorescence microscopy \cite{bednarkiewiczPhotonAvalancheLanthanide2019}. 

A key point in material development for efficient ETU is the ability to accurately model energy transfers in addition to the properties of isolated (non-interacting) ions. This is complex since ETU typically involve several rare earth ions, and thus multiple energy levels and a large number of possible energy transfers. Energy transfers are also strongly dependent on the distance between ions \cite{kushidaEnergyTransferCooperative1973b}, which means that modeling the large ensembles of ions experimentally observed will need careful averaging over ions' spatial distribution. These questions have motivated many studies, from obtaining explicit equations for the ions decay dynamics \cite{yokota1967,Inokuti1965,huberDynamicsIncoherentTransfer1981} to heavy numerical computations \cite{villanueva-delgadoSimulatingEnergyTransfer2015a}, through semi-phenomenological rate equation approaches \cite{wrightUpconversionExcitedState1976,goldnerPhotonAvalancheFluorescence1996}. 

The latter typically involve many parameters corresponding to the ions and levels involved, which have to be determined from the levels' decay rates and/or luminescence intensities. As these parameters are likely to be combined in complex ways in the decays of virtually all levels, they are difficult to evaluate from small series of measurements of a few levels and rare earth concentrations. This can make the validity of reported parameters over e.g. a broad range of dopant concentrations or excitation wavelengths, as well as their physical meaning, difficult to assess. In addition, new phenomena become important for nanostructured materials \cite{wen2018advances,Hossan2017} which are at the heart of key applications, like bio-imaging or thermometry \cite{duNanocompositesBasedLanthanidedoped2022}. It is for example well documented that in nanoparticles or thin films, defects capable of quenching luminescence are more prominent than in their bulk counterparts and are often located at free surfaces or interfaces \cite{Vetrone2004,mialonNewInsightsSize2009,chenPhotonUpconversionCore2015}. This results in further complexity in theoretical and modeling studies, especially since these defects can not always be clearly identified and/or controlled. 

In this work, we present a comprehensive study of the \Ert, \Ybt\ system in \YO\
nanoparticles. \YO\ is among the oxide crystals with the lowest phonon cut-off frequency which reduces non-radiative relaxations and levels with longer lifetimes \cite{weber1968}. \YO\ is a versatile host that can be synthesized under the forms of high-quality single bulk crystals \cite{munGrowthCharacterizationTmdoped2007}, transparent ceramics \cite{kimura2021,Brown2014} , thin films \cite{harada2020}, and nanoparticules \cite{liuDefectEngineeringQuantum2020,vetrone2003,tianUpconversionLuminescenceProperties2015a}. It finds applications in lasers \cite{sanamyanHighPowerDiodepumped2011}, luminescent probes and sensors \cite{molinaResponseCharacterizationY2O32013,chavez-garciaLuminescencePropertiesCell2021}, and, more recently, as a promising platform for optical quantum technologies \cite{kunkelRecentAdvancesRare2018a,zhongEmergingRareearthDoped2019,fossatiFrequencyMultiplexedCoherentElectrooptic2020}. All of them rely on or have to consider energy transfer phenomena. 

Our study is based on the decay times of the main transitions located between from 0.52 to 1.5 $\mu$m that have been measured as a function of \Ert\ and \Ybt\ concentrations and in singly- and co-doped samples. This was performed under direct excitation of the probed levels as well under 0.98 $\mu$m excitation in \Ybt, \Ert co-doped samples to investigate ETU process. Such an extensive approach has never been reported to our knowledge. 

Using a rate equation modeling, these data allowed us to independently extract energy transfer rates for different transitions, doping levels, and excitations wavelengths.
Luminescence quenching by defects was also observed and included in the models. This enables reproducing and predicting \Ert\ and \Ybt\ decay times for most  concentrations in singly- as well as co-doped samples and for emission excited directly or through ETU, indicating that a description in terms of populations averaged over all excited ions is adequate. Moreover, the analysis highlights a doping regime that likely optimizes ETU efficiency.
These results open the way to accurate modeling and prediction of the dynamics of RE energy level populations in complex situations. In contrast to simulations based on microscopic configurations, the rate equation modeling involves small to moderate computing power and allows for easier physical interpretation for fast, reliable, and efficient material development.

\section{Methodology}

To identify the dominant interactions between ions, we first analyze those occurring within the same species—Yb–Yb and Er–Er—before turning to Er–Yb interactions. This is achieved by examining how excited-state lifetimes, or equivalently relaxation rates, vary with dopant concentration.

Our approach combines a qualitative analysis, aimed at identifying the most likely interaction pathways for each energy level, with a quantitative analysis based on extracting relaxation rates associated with each process. This is done by modeling population dynamics using rate equations.

Rate equations can be used when all probed ions evolve similarly over time. as a function of time and can thus be accurately described by average populations for the different levels. This situation is observed for non-interacting ions, i.e. at low doping concentrations (typically $\leq 0.1-0.5$\%). In the opposite situation of high doping concentrations ($\geq 5$\%), where fast migration of energy results in identical average populations for all ions, rate equations are also appropriate. In both cases, exponential decays are experimentally observed for directly excited levels. In contrast, at intermediate concentrations, each ion is surrounded by other ions, or defects, located at varying positions. Inhomogeneous interactions lead to non-exponential decays which cannot be rigorously described by rate equations anymore.

In the following, we use rate equations across all concentrations to extract spatially averaged transfer rates, allowing for a simpler and consistent modeling approach. This avoids the complexity of spatially resolved models, while still providing useful physical insights.

Each excited-state population $n_i$ based on radiative decay rates (taken from literature measured in a high quality ceramic \cite{weber1968}), non-radiative decay rates (determined from low-concentration experiments or fitted), and energy transfer rates, which are iteratively refined throughout the study. Further details are provided in the Modeling section and in the SI.

Regarding the experimental work, time-resolved photoluminescence decays $I(t)$ of key energy levels were measured following direct excitation. For Yb-doped nanoparticles, luminescence from the $^2$F$_{5/2}$ level to the ground state was recorded, while for Er-doped nanoparticles, emissions from the $^4$S$_{3/2}$, $^4$F$_{9/2}$, $^4$I$_{11/2}$ and $^4$I$_{13/2}$ levels were measured.

We use $K_i=1/\tau_i$ as a figure of merit to compare experimental data with modeling predictions. The characteristic time $\tau_i$ is estimated from both experimental data and the model as:
\begin{equation}
\label{eq:2_tau}
    I_i(t_0+2\tau_i)=I_i(t_0)/\mathrm{e}^{2} 
\end{equation}

where $t_0$ the time of maximum intensity.

This method of estimating $\tau$ accounts for 86\% of the luminescence decay, providing a trade-off between the short- and long-lived components of the most significant portion of the emitted intensity.

\section{Modeling}

\begin{figure}[h]\centering
	\includegraphics[width=\textwidth]{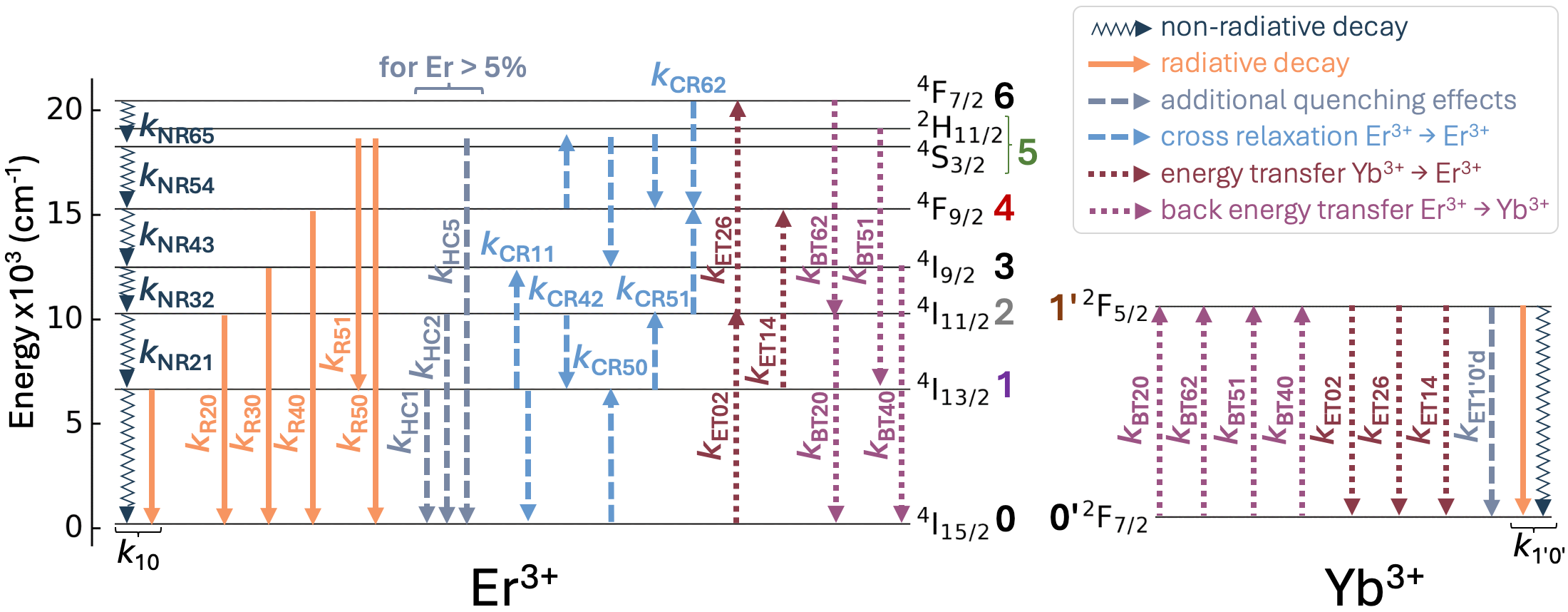}
	\caption{Energy level diagram of Yb$^{3+}$ and Er$^{3+}$ ions and illustration of the interactions considered in our modeling.}
	\label{fig:si_energylevels}
\end{figure}

We consider a system composed of Yb$^{3+}$ ions (with a concentration $N_\mathrm{Yb}$) with two energy levels, and Er$^{3+}$ ions (with a concentration $N_\mathrm{Er}$) limited to seven energy levels $i$, as presented on Figure \ref{fig:si_energylevels}. The $^2$H$_{11/2}$ and $^4$S$_{3/2}$ levels are assumed to be in thermal equilibrium and are grouped into a single level. 

In some studies on fluoride-based materials \cite{Anderson2014,Berry2015}, high up-conversion efficiency has led to the observation of blue emissions originating from three-photon processes. Although such effects have also been reported in \YO\ systems \cite{Song2004}, in our nanoparticles—excited at low excitation power—no signatures of three-photon processes were observed. This confirms that the observed emissions result solely from two-photon up-conversion, with no additional excited states involved.

To model ion interactions, we employ a system of rate equations describing the time evolution of the population density 
$n_i$ of each excited-state level $i$, according to the relation $dn_i/dt$, as a function of the interaction rates 
$k$ and the populations of other relevant energy levels \cite{auzel1973}. This approach assumes rapid energy migration and relies on average decay rates across the system. The following processes are included in the model:
\begin{enumerate}
    \item radiative decays $k_{\mathrm{R}ij}$ from a level $i$ to a level $j$
    
    \item non-radiative decays $k_{\mathrm{NR}i,i-1}$ from a level $i$ to the level $i-1$ caused by interactions with lattice phonons
    
    \item energy transfer Yb$\rightarrow$Er $k_{\mathrm{ET}ij}$ populating a level $j$ from a level $i$ of Er$^{3+}$ ions
    
    \item back energy transfer (BT) Er$\rightarrow$Yb $k_{\mathrm{BT}ij}$ populating a level $j$ from a level $i$ of Er$^{3+}$ ions
    \item  cross-relaxations (CR) Er$\rightarrow$Er $k_{\mathrm{CR}ij}$ depopulating the levels $i$ and $j$ of Er$^{3+}$ ions

    \item additional defect-related quenching path $k_{\mathrm{ET1'0'd}}$ depopulating the \Ybt\ excited state level

    \item additional defect-related quenching paths observed for high doping concentrations (> 5\%) $k_{\mathrm{HC}i}$ depopulating $i$ level of \Ert\ ions
\end{enumerate}

As mentioned in the Methodology, radiative decay rates are taken from literature \cite{weber1968,sardar2007}. Non-radiative decay rates are either determined from low-concentration experiments, fitted to experimental data, or estimated based on the energy gap. Some  energy transfer rates—including BT and CR processes —are iteratively refined throughout the study. The remaining rates are fixed depending on whether the process is resonant or non-resonant (phonon-assisted) \cite{yamada1972}.

The mechanisms taken into account for this modeling are represented on Figure \ref{fig:si_energylevels}. Further details are provided in the Supplementary Information.

\section{Experimental}

\subsection{Synthesis of Yb$^{3+}$ and Er$^{3+}$ doped Y$_2$O$_3$ nanoparticles}

Crystalline Y$_2$O$_3$-based nanoparticles doped with Yb$^{3+}$ and Er$^{3+}$ ions were successfully synthesized through a homogeneous precipitation route followed by carefully controlled thermal treatments \cite{liu2018controlled,deoliveiralimaInfluenceDefectsSubA2015}. In brief, Y(OH)CO$_3$·nH$_2$O was employed as the precursor material. The homogeneous precipitation was induced via the thermolysis of urea in an aqueous medium containing Y(NO$_3$)$_3$·6H$_2$O (99.99\% purity, Alfa Aesar®) and urea (99\% purity, Sigma-Aldrich®), at final concentrations of 7.5 mmol·L$^{-1}$ and 3 mol·L$^{-1}$, respectively. The dopant ions Er$^{3+}$ and Yb$^{3+}$ were introduced via their respective nitrate salts, Er(NO$_3$)$_3$·6H$_2$O and Yb(NO$_3$)$_3$·5H$_2$O, with purities of 99.99\% and 99.9999\% (Alfa Aesar®).

A total of ten samples were prepared with varying dopant concentrations. For Er$^{3+}$-doped nanoparticles, the molar doping relative to Y$^{3+}$ were 0.5\%, 2\%, 7\%, and 17\%. In the case of Yb$^{3+}$ doping, the compositions included 0.5\%, 1\%, 2\%, 5\%, 9\%, and 17\% molar ratios with respect to Y$^{3+}$ molar amount.

Further details about the synthesis are provided in Supplementary Information.

\subsection{Pulsed optical parametric oscillator}

For PL spectra and decays, Yb$^{3+}$ and Er$^{3+}$ ions were excited by 6 ns long pulses from a tunable (linewidth <5 cm$^{-1}$ ) optical parametric oscillator (OPO) pumped by a Nd:YAG Q-switched laser (Ekspla NT342B-SH). Visible fluorescence was collected and detected through a spectrometer (Acton SP2300) equipped with three different gratings and an intensiﬁed charge-coupled device (ICCD) camera (Princeton Instruments PI-MAX4) for spectra or a photo-multiplier tube (S20 cathode) for decays. NIR spectra and decays were measured with an InGaAs photodiode detector (FEMTO OE-200).

\section{Results and discussion}

\subsection{Morphology and emission spectra}

\begin{figure}[h]\centering
	\includegraphics[width=\textwidth]{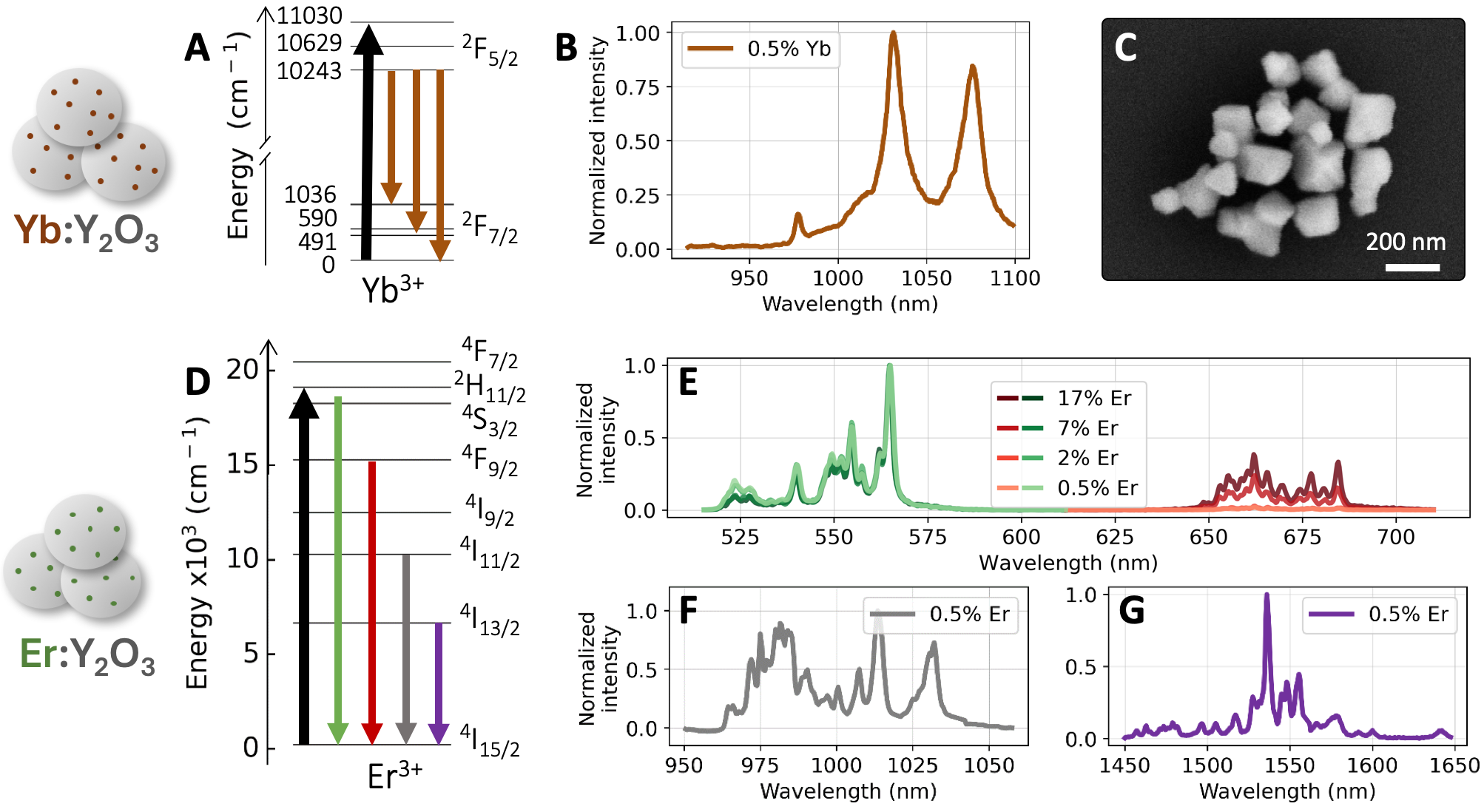}
	\caption{\textbf{Singly-doped nanoparticles - Emission spectra and morphology.} (A) Energy level diagram of \Ybt, highlighting the three main emission transitions corresponding to peaks at 977 nm, 1030 nm, and 1075 nm. (B) Emission spectra of nanoparticles doped with 0.5\% Yb, showing the emission lines between the Stark levels associated with the transitions in (A). (C) SEM image of nanoparticles doped with 0.5\% Yb, revealing a spherical morphology and an average diameter of approximately 100 nm. (D) Energy level diagram of \Ert. (E) Emission spectra in the visible range following excitation at 520 nm (indicated by the black arrow), normalized to the green emission peak. A decrease in the red-to-green emission ratio is observed with increasing Er$^{3+}$ doping concentration. (F,G) Emission spectra in the near-infrared (1 $\mu$m) range (corresponding to the $^4$I$_{11/2}$$\rightarrow$$^4$I$_{15/2}$ transition) and around 1.5 $\mu$m (corresponding to the $^4$I$_{13/2}$$\rightarrow$$^4$I$_{15/2}$ transition) of nanoparticles doped with 0.5\% Er$^{3+}$ excited at 520 nm.}
	\label{fig:panel1}
\end{figure}

Nanoparticles morphology was investigated using Scanning Electron Microscopy (SEM). Figure \ref{fig:panel1}.C shows SEM images of nanoparticles doped with 0.5\% Yb, confirming their spherical morphology and approximate size of 100 nm. This synthesis procedure leads to single-crystalline nanoparticles in \YO\ cubic phase (I-3a) with limited aggregation \cite{liuDefectEngineeringQuantum2020}. \Ybt\ and \Ert\ ions substitute Y$^{3+}$ ions at both C$_2$ (24-fold multiplicity) and C$_\mathrm{3i}$ (8-fold multiplicity) crystallographic sites in the \YO\ lattice. Due to the presence of an inversion center at C$_\mathrm{3i}$ sites, electric dipole (ED) transitions are forbidden for ions occupying these positions \cite{peacock2007intensities}. However, magnetic dipole (MD) transitions, governed by the selection rule $\Delta J = 0, \pm1$, remain allowed and can exhibit non-zero intensity. As a result, only the $^2$F$_{5/2}$ $\to$ $^2$F$_{7/2}$ transition of \Ybt\ ions and the $^4$I$_{13/2}$ $\to$ $^4$I$_{15/2}$ transition of \Ert\ ions are observed from C$_\mathrm{3i}$ sites.
Thus, most of the emissions analyzed in this study originate from transitions of ions located at C$_2$ sites.

To evaluate the luminescent properties of the rare-earth ions, emission spectra of Yb-doped nanoparticles were recorded at room-temperature, by exciting the $^4$F$_{7/2}$ $\rightarrow$ $^4$F$_{5/2}$ transition at 905 nm. The near-infrared (NIR) emission spectrum presented on Figure \ref{fig:panel1}.B, displays three main peaks at 977 nm, 1030 nm and 1075 nm, attributed to transitions between the Stark levels of $^2$F$_{5/2}$ and $^2$F$_{7/2}$ manifolds, as shown in Figure \ref{fig:panel1}.A \cite{chang1982,gruber1985}. Moreover, the recorded spectra are consistent with those expected for Y$_2$O$_3$, confirming the good crystallization of the material. No significant spectral variations are observed across the different Yb$^{3+}$ doping concentrations.

The NIR emission spectra of \Ert\ doped nanoparticles are presented in Figure \ref{fig:panel1}.F and Figure \ref{fig:panel1}.G, after exciting the $^4$I$_{15/2}$ $\rightarrow$ $^2$H$_{11/2}$ transition at 520 nm. Emission around 1 $\mu$m is attributed to the $^4$I$_{11/2}$ $\rightarrow$ $^4$I$_{15/2}$ transition, while the emission around 1.5 $\mu$m corresponds to the $^4$I$_{13/2}$ $\rightarrow$ $^4$I$_{15/2}$ transition \cite{Brown2014,Li2016}.
Finally, the visible emission spectra of Er-doped nanoparticles, following excitation at 520 nm, are shown in Figure \ref{fig:panel1}.E. The green emission around 550 nm is attributed to relaxation from the thermally coupled $^2$H$_{11/2}$ and $^4$S$_{3/2}$ levels to the ground state. On the other hand, the red emission around 660 nm arises from the $^4$F$_{9/2}$ $\rightarrow$ $^4$I$_{15/2}$ transition. An increase in red emission relative to green emission is observed with increasing Er$^{3+}$ concentration, as previously reported \cite{capobianco2002}. This enhancement is attributed to cross-relaxation mechanisms, which result in populating more efficiently the $^4$F$_{9/2}$ level at higher concentrations. A more detailed discussion of these mechanisms is provided later.

\subsection{Singly-doped nanoparticles}

\begin{figure}[h!]\centering
	\includegraphics[width=0.85\textwidth]{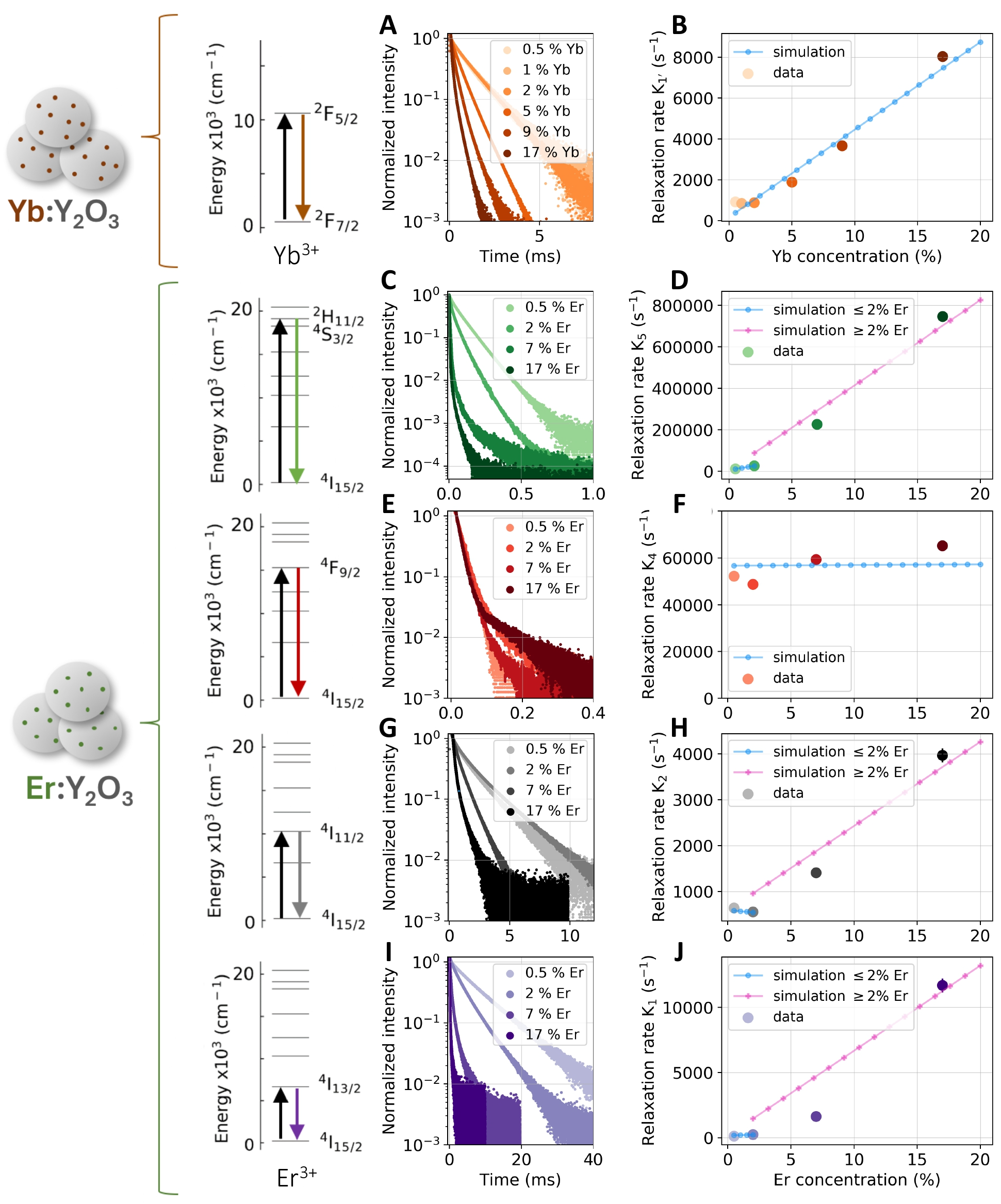}
	\caption{\textbf{Singly-doped nanoparticles - under direct excitation.} (A) Luminescence decay of Yb$^{3+}$ excited state level for various doping levels and (B) variation of the relaxation rate $K_{1'}$ as a function of Yb$^{3+}$ concentration. (C) Luminescence decay of Er$^{3+}$ from the green-emitting levels $^2$H$_{11/2}$,$^4$S$_{3/2}$$\to$\qIq\ and (D) variation of the relaxation rate $K_5$ as a function of Er$^{3+}$ concentration. (E) Luminescence decay of Er$^{3+}$ red-emitting level $^4$F$_{9/2}$$\to$\qIq\ and (F) variation of the relaxation rate $K_4$ as a function of Er$^{3+}$ concentration. (G) Luminescence decay of Er$^{3+}$ around 1 $\mu$m $^4$I$_{11/2}$$\to$\qIq\ and (H) variation of the relaxation rate $K_2$ as a function of Er$^{3+}$ concentration. (I) Luminescence decay of Er$^{3+}$ around 1.5 $\mu$m $^4$I$_{13/2}$$\to$\qIq\ and (J) variation of the relaxation rate $K_1$ as a function of Er$^{3+}$ concentration.}
	\label{fig:panel2}
\end{figure}

To gain both qualitative and quantitative insight into the mechanisms occurring between ions of the same species—\Ybt-\Ybt\ and \Ert-\Ert— such as non-radiative decays, cross-relaxations (CR) and additional defect-related quenching, we first investigated singly doped nanoparticles. Photoluminescence decays $I(t)$ of key energy levels were thus measured under direct excitation and are presented in Figure \ref{fig:panel2}. Table \ref{tab:chap5:tau_yb_np} (resp. Table \ref{tab:chap5:tau_er_np}) summarizes up experimental lifetimes obtained in Yb- (resp. Er-) doped nanoparticles.

\begin{table}[H]
\begin{centering}
\begin{tabular}{c|cccccc|c}
& \multicolumn{7}{c}{Yb concentration}\\ \cline{2-8}
                      & \multicolumn{6}{c|}{\textit{nanoparticles}} & \textit{ceramic} \cite{Zhang2015} 
\\
   \textit{level}  & 0.5 \% & 1 \% & 2 \%   & 5 \%  & 9 \%  & 17 \%   &1 \%    
\\ \hline
   $^2$F$_{5/2}$           & 1.1& 1.2& 1.1& 0.53&0.27& 0.12& 0.91  
     \end{tabular}
\caption{\Ybt\ excited state lifetimes (ms) measured for different doping concentrations in the singly-doped nanoparticles, and compared with the one reported for a ceramic sample \cite{Zhang2015}.}
\label{tab:chap5:tau_yb_np}
\end{centering}
\end{table}

\begin{table}[H]
\begin{centering}
\begin{tabular}{c|cccc|c}
& \multicolumn{5}{c}{Er concentration}\\ \cline{2-6}
                       & \multicolumn{4}{c|}{\textit{nanoparticles}} & \textit{ceramic} \cite{weber1968} 
\\
    \textit{level}& 0.5 \%  & 2 \%   & 7 \%   & 17 \%   &0.5 \%    
\\  \hline 
   \dHo,\qSt\           & 0.085   & 0.038  & 0.004  & 0.001   &0.088     
\\
                                            \qFn\          & 0.019   & 0.021  & 0.017  & 0.015   &0.030     
\\
                                            $^4$I$_{11/2}$         & 1.56     & 1.80    & 0.71   & 0.25    &3.9       
\\
                                           $^4$I$_{13/2}$           & 8.1       & 4.0    & 0.6    & 0.09   
 &8.0       \end{tabular}
\caption{\Ert\ lifetimes (ms) of different energy levels for various doping concentrations of the singly-doped nanoparticles, and compared to values reported for a ceramic sample \cite{weber1968}.}
\label{tab:chap5:tau_er_np}
\end{centering}
\end{table}

\subsubsection{Yb$^{3+}$ doped nanoparticles}

Figure \ref{fig:panel2}.A presents Yb$^{3+}$ luminescence decay curves following pulsed excitation at 905 nm. At low concentration (up to 2\%), the $^2$F$_{5/2}$ lifetime $\tau$ remains around 1 ms consistent with previous studies in ceramics, confirming the high crystalline quality of the nanoparticles \cite{kong2004,Zhang2015}. For higher doped nanoparticles (5\%, 9\% and 17\%),  the decay accelerates significantly, with $\tau$ decreasing to approximately 0.12 ms at the highest Yb$^{3+}$ concentration. This trend suggests an increased probability of relaxation through non-radiative mechanisms, likely driven by energy diffusion among Yb$^{3+}$ ions, followed by energy transfer to centers that relax non-radiatively \cite{Auzel2003}. 

Furthermore, for this range of high concentrations, the fluorescence decay does not follow a purely mono-exponential function over time. Instead, an initial rapid intensity drop is observed, which then transitions into a mono-exponential behavior. This characteristic trend is indicative of a distribution of energy transfers \Ybt\ to defects, arising from variations in the local environment of individual donors. As a result, the interaction strengths among Yb$^{3+}$ ions differ, leading to a distribution of lifetimes \cite{Weber1971,Wang2011}. The experimental decay curves $I(t)$ are well-fitted using the model developed by Inokuti and Hirayama for dipole-dipole interactions between donors and acceptors, with an interaction parameter of $s = 6$ \cite{Inokuti1965} (Figure S\ref{fig:si_yb_inokuti}). 

Figure \ref{fig:panel2}.B presents the relaxation rate $K_{1'}$, estimated using Equation \ref{eq:2_tau}, as a function of Yb$^{3+}$ concentration. We observe two regimes: at low concentrations up to 2\%, $K_{1'}$ is nearly constant and from 5\% it increases linearly.

The first regime is attributed to limited energy migration, where only Yb$^{3+}$ ions in close proximity to defects undergo quenching. In the second regime, as Yb$^{3+}$ concentration increases, a larger fraction of Yb$^{3+}$ ions are situated near defects, due to increased energy migration, leading to enhanced quenching. This manifests by a rapid initial decay followed by an exponential tail. This exponential component can be described by an additional relaxation term of the form $k = CN_AN_D$, where $N_A$ represents the concentration of acceptors (i.e., defects), $N_D$ corresponds to the concentration of donors (i.e., Yb$^{3+}$ ions), and $C$ is a parameter accounting for both diffusion among the donors and energy transfer to the acceptors \cite{Weber1971}. The observed linear dependence suggests that the defect concentration remains constant across the studied range. In the following, we note $k_{\mathrm{ET1'0'd}}$ the energy transfer rate from Yb$^{3+}$ ions to defects where $k_{\mathrm{ET1'0'd}}=CN_A$, to incorporate this concentration quenching into the modeling.

Consequently, we express the relaxation rate $K_{1'}$ of the \Ybt\ excited state as $K_{1'} = k_{1'0'} + k_{\mathrm{ET1'0'd}} n_{0'}$, where $k_{1'0'}$ accounts for both radiative and non-radiative decay rates, and $n_0'$ $\approx$ $N_{\mathrm{Yb}}$ assuming low excitation power. This approach enables accurate modeling across a broad concentration range. A linear fit of $K_{1'}=f(N_{\mathrm{Yb}})$ leads to $k_{1'0'}=173$ s$^{-1}$ and $k_{\mathrm{ET1'0'd}}=2.2\times10^{-18}$ cm$^3$.s$^{-1}$. The relaxation rate values, obtained from the simulation over a wide concentration range, are plotted in blue in Figure \ref{fig:panel2}.B. These values exhibit good agreement with the experimental data in the concentration range of 2 to 17\%, which is the primary focus of this study. The relaxation rate is slightly underestimated in the 0.5 to 1\% range; however, this lower concentration range is not relevant to the scope of our work.

\vspace{0.5cm}

In summary, the analysis reveals a clear trend of increased luminescence decay with higher Yb$^{3+}$ concentration, particularly due to energy transfer to defects. By adding and adjusting the energy transfer rate $k_{\mathrm{ET1'0'd}}$, the model developed effectively describes these interactions.

\newpage

\subsubsection{Er$^{3+}$ doped nanoparticles}

Er$^{3+}$ ions possess a rich energy level structure with multiple excited states involved in energy transfer and relaxation processes. To identify the dominant interactions contributing to ETU, we focus on four key levels that play central roles in the emission dynamics: $^2$H$_{11/2}$,$^4$S$_{3/2}$ (green emission), $^4$F$_{9/2}$ (red emission), $^4$I$_{11/2}$ (1$\mu$m emission), $^4$I$_{13/2}$ (1.5 $\mu$m emission).

\textbf{Green emission.} Figure \ref{fig:panel2}.C shows the luminescence decay curves corresponding to transitions from the thermalized $^2$H$_{11/2}$ and $^4$S$_{3/2}$ levels to the ground state in nanoparticles doped with 0.5\%, 2\%, 7\%, and 17\% of \Ert. 

The lifetime of the green-emitting levels in nanoparticles doped with 0.5\% Er$^{3+}$ is estimated to be 85 $\mu$s. This value is consistent with those reported for high-quality reference materials, such as 88 $\mu$s in ceramics containing 0.2\% Er$^{3+}$ \cite{weber1968}, and approximately 100 $\mu$s in high-quality polycrystalline thin films with Er$^{3+}$ concentrations below 1\% \cite{balasa2024}. The luminescence decay curves deviate from a mono-exponential trend, likely due to inhomogeneities in energy transfer probabilities, similar to what is observed for Yb$^{3+}$ ions. Moreover, the decays reveal a pronounced decrease in lifetime with increasing Er$^{3+}$ concentration. This effect is attributed to the cross-relaxation process CR50 ($^2$H$_{11/2}$,$^4$I$_{15/2}$)$\rightarrow$($^4$I$_{9/2}$,$^4$I$_{13/2}$) \cite{balasa2024,auzel1973}. This non-resonant process characterized by an energy mismatch of $\Delta E = 185$ cm$^{-1}$, is highly efficient due to 1) its involvement of the ground state, which remains predominantly populated, and 2) its energy gap being significantly smaller than the maximum phonon energy in the crystal ($\approx$ 550 cm$^{-1}$).

Figure \ref{fig:panel2}.D illustrates the variation of the relaxation rate $K_5$ as a function of Er$^{3+}$ concentration. CR50 mechanism contributes to the relaxation rate $K_5$ as $k_\mathrm{CR50}n_0$, where $k_{\mathrm{CR50}}$ represents the associated energy transfer rate, and $n_0$ $\approx$ $N_{\mathrm{Er}}$ assuming low excitation power. Thus, if CR50 interaction was the dominant contributor, $K_5$ would exhibit a linear dependence on the dopant concentration. However, deviations from this trend at high doping levels indicate the presence of additional quenching mechanisms, likely related to surface defects. To account for this, we distinguish between two regimes in our modeling.

For low Er$^{3+}$ concentrations ($\leq$2\%), the relaxation rate is well described by a linear model: $K_5 = 
 k_{\mathrm{R}50}+k_{\mathrm{R}51}+k_{\mathrm{NR}54}+k_{\mathrm{CR50}} n_{0}$. In this regime, a linear fit—with the radiative rates $k_{\mathrm{R}50}$ and $k_{\mathrm{R}51}$ fixed (see Supplementary Information)—leads to $k_{\mathrm{NR}54}=5.5\times10^3$ s$^{-1}$ and $k_{\mathrm{CR50}}=4.9\times10^{-17}$ cm$^3$.s$^{-1}$. 
 
 At higher Er$^{3+}$ concentrations ($\geq$2\%), an additional quenching term $k_{5\mathrm{HC}}$ is introduced to account for enhanced quenching effects. The modified expression becomes: $K_5 = k_{\mathrm{R}50}+k_{\mathrm{R}51}+k_{\mathrm{NR}54} + k_{\mathrm{CR50}} n_{0} + k_{5\mathrm{HC}} n_{0}$. Fitting the data with only $k_{5\mathrm{HC}}$ as a free parameter, gives $k_{5\mathrm{HC}}=1.6\times10^{-16}$ cm$^3$.s$^{-1}$. 
 
 The relaxation rates extracted from simulations—plotted in blue for the low-concentration regime and in pink for the high-concentration case in Figure \ref{fig:panel2}.D, show good agreement with experimental data.

\textbf{Red emission.} Figure \ref{fig:panel2}.E presents the luminescence decay curves from the red-emitting level corresponding to the transition $^4$F$_{9/2}$ $\rightarrow$ $^4$I$_{15/2}$ in nanoparticles doped with different concentrations of \Ert. 

The luminescence decays exhibit a nearly mono-exponential behavior and remain largely unaffected by increasing dopant concentration. As shown in Figure \ref{fig:panel2}.F, the relaxation rate $K_4$ varies between $4.9\times10^4$ s$^{-1}$ and $6.5\times10^4$ s$^{-1}$. This high relaxation rate is primarily attributed to the small energy gap to the adjacent $^4$I$_{9/2}$ level, which enables an efficient multiphonon relaxation pathway.

A linear fit of $K_{4}=k_{\mathrm{R}40}+k_{\mathrm{NR}43}$ leads to $k_{\mathrm{NR}43}=5.5\times10^4$ s$^{-1}$, with $k_{\mathrm{R}40}$ fixed to a literature value.
The relaxation rates estimated from the simulation, plotted in blue in Figure \ref{fig:panel2}.F, are close to the experimental data.

\textbf{Red-to-green ratio.} The increase in the red-to-green (R/G) ratio with Er$^{3+}$ concentration can be attributed to the role of the CR50 process, which populates the $^4$I$_{9/2}$ and $^4$I$_{13/2}$ levels. The $^4$I$_{9/2}$ level feeds efficiently to $^4$I$_{11/2}$ via multiphonon relaxation due to the relatively small energy gap ($\approx$ 2215 cm$^{-1}$). Consequently, the ETU process ($^4$I$_{11/2}$, $^4$I$_{13/2}$) $\rightarrow$ ($^4$F$_{9/2}$, $^4$I$_{15/2}$) becomes more efficient at higher Er$^{3+}$ concentrations, leading to the observed increase in the R/G ratio.

\textbf{Long time-scale component in visible decays induced by energy transfer from NIR levels.}

For highly doped nanoparticles, a rapid initial luminescence decay is followed by a slower decay component in the green-emitting levels ($N_{\mathrm{Er}}$ > 7\%, after approximately 150 $\mu$s) and red-emitting levels ($N_{\mathrm{Er}}$ > 2\%, after approximately 0.1 ms), characteristic of an ET process \cite{vetrone2003}. The slower decay in the green-emitting levels results from a ($^4$I$_{11/2}$, $^4$I$_{11/2}$) $\rightarrow$ ($^4$F$_{7/2}$, $^4$I$_{15/2}$) process. The $^4$I$_{11/2}$ level is populated via multiphonon relaxations, following a cascade from the $^4$S$_{3/2}$ level to $^4$F$_{9/2}$, then to $^4$I$_{9/2}$, and finally to $^4$I$_{11/2}$. Once the $^4$F$_{7/2}$ level is populated via up conversion, the $^2$H$_{11/2}$ and $^4$S$_{3/2}$ levels are replenished through multiphonon relaxation. Given the extended lifetime of the $^4$I$_{11/2}$ level, it serves as a reservoir, feeding the green-emitting levels and resulting in the long-tail behavior observed in the fluorescence decay.

For the red-emitting level, the $^4$F$_{9/2}$ level is populated via two distinct pathways. The first involves multiphonon relaxation from the $^4$S$_{3/2}$ level, which itself is fed by ETU. The second occurs through the energy transfer process ($^4$I$_{11/2}$, $^4$I$_{13/2}$) $\rightarrow$ ($^4$F$_{9/2}$, $^4$I$_{15/2}$). 

These mechanisms are considered negligible and are not included in our model, as the \Ybt$\to$\Ert\ ETU process is significantly more efficient in the dynamics of co-doped nanoparticles.

\textbf{1 $\mu$m emission.} The luminescence decays corresponding to the $^4$I$_{11/2}$ $\rightarrow$ $^4$I$_{15/2}$ transition in \Ert\ doped nanoparticles are presented in Figure \ref{fig:panel2}.G.

The decay time measured on the 2\% doped 
nanoparticles is slightly longer (1.8 ms) than the one obtained for 0.5\% Er$^{3+}$ nanoparticles (1.6 ms). This could be attributed to reabsorption effects, also known as radiation trapping \cite{Auzel2003}, or to the up-conversion mechanism CR11 ($^4$I$_{13/2}$, $^4$I$_{13/2}$) $\rightarrow$ ($^4$I$_{9/2}$, $^4$I$_{15/2}$) \cite{Hossan2017}, which may repopulate $^4$I$_{11/2}$ via multiphonon relaxation from the $^4$I$_{9/2}$ level, with the long-lived $^4$I$_{13/2}$ level serving as a reservoir. This non-resonant process, characterized by a energy mismatch of 628 cm$^{-1}$, could lengthen the lifetime of the $^4$I$_{11/2}$ level.

For higher doping levels, the decay times are significantly shortened, indicating important concentration quenching. This could be due to the up-conversion mechanism CR22 ($^4$I$_{11/2}$, $^4$I$_{11/2}$) $\rightarrow$ ($^4$F$_{7/2}$, $^4$I$_{15/2}$) as introduced earlier. This mechanism depends quadratically on the population of the $^4$I$_{11/2}$ level, so a change in laser power should lead to a significant difference in the decay time. However, when the excitation power was experimentally reduced by a factor of 50, only small changes were observed, indicating that the significant concentration quenching is more likely due to migration and energy transfer to defects.

Figure \ref{fig:panel2}.H shows the evolution of the relaxation rates $K_2$ as a function of Er$^{3+}$ concentration. Although the number of data points is limited, the observed trend is similar to that of the Yb$^{3+}$ transition. At lower concentrations, the quenching remains minimal, indicating a regime of limited diffusion. As the Er$^{3+}$ concentration increases, a linear trend emerges, suggesting enhanced energy migration towards defects. 

Once again, the modeling can be separated into two distinct regimes. For low Er$^{3+}$ concentrations ($\leq$2\%), no additional quenching processes are considered, and the relaxation rate is described by: $K_2=k_{\mathrm{R}20}+k_{\mathrm{NR}21}$. A linear fit yields $k_{\mathrm{NR}21}=448$ s$^{-1}$, where $k_{\mathrm{R}20}$ value was fixed. For higher Er$^{3+}$ concentrations ($\geq$2\%), an additional quenching term $k_{2\mathrm{HC}}$ is introduced to account for increased energy transfer processes, leading to: $K_2=k_{\mathrm{R}20}+k_{\mathrm{NR}21}+k_{2\mathrm{HC}}n_0$. A fit, with only $k_{2\mathrm{HC}}$ as a free parameter, gives $k_{2\mathrm{HC}}=9.4\times10^{-19}$ cm$^3$.s$^{-1}$.

The relaxation rates $K_2$ obtained from simulations are plotted in blue (low Er$^{3+}$ regime) and pink (high Er$^{3+}$ regime) in Figure \ref{fig:panel2}.H. The model reliably reproduces the measured relaxation rates, particularly at low doping levels. Additionally, a slight decrease in the simulated relaxation rate is observed at low concentrations, primarily due to the inclusion of the CR11 mechanism in the simulation, since radiation trapping effects are not included in the modeling.

\textbf{1.5 $\mu$m emission.} Figure \ref{fig:panel2}.I shows the luminescence decays corresponding to the transition $^4$I$_{13/2}$ $\rightarrow$ $^4$I$_{15/2}$ in Er-doped nanoparticles. 

The luminescence decay measurements once again reveal a significant reduction in lifetime with increasing Er$^{3+}$ concentration. For a doping concentration of 0.5\%, the measured lifetime is 8.1 ms, whereas it decreases to 0.09 ms for nanoparticles doped with 17\%.

Moreover, while the 0.5 \% Er-doped nanoparticles follows a mono-exponential trend, higher doping levels introduce a fast-decay component before stabilizing into a more mono-exponential behavior, as previously observed.

This quenching effect could be attributed to the up-conversion process CR11, discussed earlier. However, as with the $^4$I$_{11/2}$ level, no significant lifetime variation was observed upon reducing the excitation power, whereas a strong dependence would be expected if CR11 were the dominant quenching mechanism. This suggests that quenching primarily results from energy migration and transfer to defects.

Figure \ref{fig:panel2}.J shows the evolution of the relaxation rates $K_1$ as a function of Er$^{3+}$ concentration. Once again, the behavior follows a similar trend to the Yb$^{3+}$ transition and the $^4$I$_{11/2}$ $\rightarrow$ $^4$I$_{15/2}$ Er$^{3+}$ transition. For the first regime of low Er$^{3+}$ concentration ($\leq$2\%), $k_{10}$—including both radiative and non-radiative rates— is adjusted by fitting $K_1$ following the relation $K_1=k_{10}$, yielding $k_{10}=186$ s$^{-1}$. For the higher Er$^{3+}$ concentration case, $k_{1\mathrm{HC}}$ is introduced to take into account the quenching such as: $K_1=k_{10}+k_{1\mathrm{HC}}n_0$. A fit gives $k_{1\mathrm{HC}}=3.1\times10^{-18}$ cm$^3$.s$^{-1}$. The relaxation rates $K_1$ obtained from simulations, plotted in blue (low Er$^{3+}$ concentration) and in pink (high Er$^{3+}$ concentration) in Figure \ref{fig:panel2}.J, accurately reproduces the experimental data, particularly at low doping levels.  In this regime, a slight increase in the simulated relaxation rate with higher Er$^{3+}$ concentrations is observed, which is attributed to the CR11 mechanism. The corresponding energy transfer rate was estimated by considering the energy gap ($\Delta E =628$ cm$^{-1}$), and its minimal impact on the relaxation rate further supports that quenching is primarily driven by other mechanisms, likely defect-related.

\vspace{0.5cm}

To summarize, the luminescence dynamics of Er-doped nanoparticles reveal strong concen-tration-dependent quenching, particularly affecting green and near-infrared emissions. At low Er$^{3+}$ levels ($\leq$2\%), the decay behavior aligns with high-quality reference materials and is well-captured by linear models. At higher concentrations, additional quenching mechanisms—such as cross-relaxation and energy migration to defects—significantly shorten lifetimes. ETU among \Ert\ ions contribute to long-tail decays. Overall, the simulated relaxation rates show good agreement with experimental data. The rate of the various relaxation paths extracted from experimental data are summarized in Table~\ref{tab3}.

\begin{table}[H]
\begin{centering}

\begin{tabular}{c|c||c|c|c|c}
\multicolumn{1}{c|}{ion}                      & \multicolumn{1}{c||}{level}                                   & \begin{tabular}[c]{@{}c@{}}*radiative \\ rate\end{tabular}                               & \begin{tabular}[c]{@{}c@{}}non-radiative \\ rate\end{tabular} & \begin{tabular}[c]{@{}c@{}}additional \\ relaxation paths\end{tabular}                                    & \begin{tabular}[c]{@{}c@{}}additional \\ relaxation paths\\ for concentration \textgreater 2\%\end{tabular} \\
                                          &                                                         & (s$^{-1}$)                                                                                        & (s$^{-1}$)                                                               & (cm$^3$.s$^{-1}$)                                                                                         & (cm$^3$.s$^{-1}$)                                                                                           \\ \hline \hline
\multicolumn{1}{c|}{\textbf{Yb$^{3+}$}}                   & $^2$F$_{5/2}$                                           & \multicolumn{2}{c|}{\begin{tabular}[c]{@{}c@{}}$k_{1'0'}$\\ 173\end{tabular}}                                                                                                 & \begin{tabular}[c]{@{}c@{}}$k_{\mathrm{ET1'0'd}}$\\ $2.2\times10^{-18}$\end{tabular}                      & /                                                                                  \\ \hline 
\multicolumn{1}{c|}{}                     & \dHo,\qSt\ & \begin{tabular}[c]{@{}c@{}}*$k_{\mathrm{R}50}$\\ $9.8\times10^{2}$\\ *$k_{\mathrm{R}51}$\\ $4.2\times10^2$\end{tabular} & \begin{tabular}[c]{@{}c@{}}$k_{\mathrm{NR}54}$\\ $5.5\times10^{3}$\end{tabular}     & \begin{tabular}[c]{@{}c@{}}$k_{\mathrm{CR50}}$\\ $4.9\times10^{-17}$\end{tabular} & \begin{tabular}[c]{@{}c@{}}$k_{5\mathrm{HC}}$\\ $1.6\times10^{-16}$\end{tabular}             \\ \cline{2-6} 
\multicolumn{1}{c|}{}                     & $^4$F$_{9/2}$                                           & \begin{tabular}[c]{@{}c@{}}*$k_{\mathrm{R}40}$\\ $1.7\times10^3$\end{tabular}                                & \begin{tabular}[c]{@{}c@{}}$k_{\mathrm{NR}43}$\\ $5.5\times10^4$\end{tabular}       & /                                                                               & /                                                                                  \\ \cline{2-6} 
\multicolumn{1}{c|}{}                     & $^4$I$_{11/2}$                                          & \begin{tabular}[c]{@{}c@{}}*$k_{\mathrm{R}20}$\\ $1.5\times10^2$\end{tabular}                                & \begin{tabular}[c]{@{}c@{}}$k_{\mathrm{NR}21}$\\ 448\end{tabular}                   & /                                                                                 & \begin{tabular}[c]{@{}c@{}}$k_{2\mathrm{HC}}$\\ $9.4\times10^{-19}$\end{tabular}             \\ \cline{2-6} 
\multicolumn{1}{c|}{\multirow{-8}{*}{\textbf{Er$^{3+}$}}} & $^4$I$_{13/2}$                                          & \multicolumn{2}{c|}{\begin{tabular}[c]{@{}c@{}}$k_{10}$\\ 186\end{tabular}}                                                                                                   & /                                                                                & \begin{tabular}[c]{@{}c@{}}$k_{1\mathrm{HC}}$\\ $3.1\times10^{-18}$\end{tabular}            
\end{tabular}
\caption{Relaxation rates of Yb$^{3+}$ and Er$^{3+}$ excited states determined by fitting experimental data. *Radiative rates are taken from literature \cite{weber1968}.}
\label{tab3}
\end{centering}
\end{table}

\newpage

\subsection{Co-doped nanoparticles}

\begin{figure}[h!]\centering
	\includegraphics[width=\textwidth]{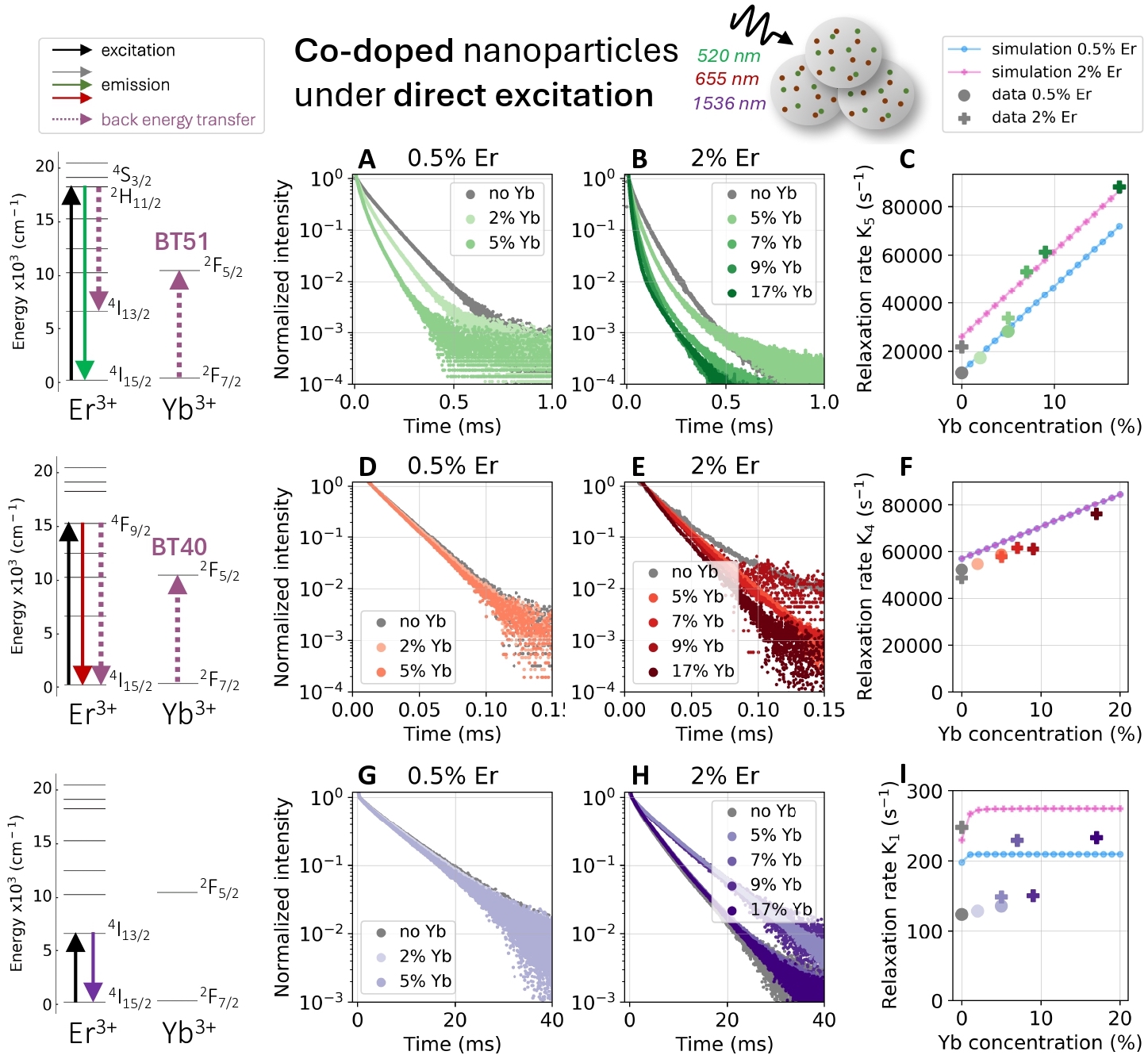}
	\caption{\textbf{Co-doped nanoparticles - under direct excitation.} (A,B) Luminescence decays of Er$^{3+}$ green-emitting levels ($^2$H$_{11/2}$,$^4$S$_{3/2}$) of nanoparticles co-doped with 0.5 \% and 2 \% of Er. (C) Variation of the relaxation rate $K_5$ as a function of Yb$^{3+}$ doping concentration. (D,E) Luminescence decays of Er$^{3+}$ red-emitting level ($^4$F$_{9/2}$) of nanoparticles co-doped with 0.5 \% and 2 \% of Er. (F) Variation of the relaxation rate $K_4$ as a function of Yb$^{3+}$ concentration. (G,H) Luminescence decays of Er$^{3+}$ 1.5 $\mu$m-emitting level ($^4$I$_{13/2}$) of nanoparticles co-doped with 0.5 \% and 2 \% Er. (I) Variation of the relaxation rate $K_1$ as a function of Yb$^{3+}$ concentration.}
	\label{fig:panel3}
\end{figure}

\begin{figure}[h!]\centering
	\includegraphics[width=\textwidth]{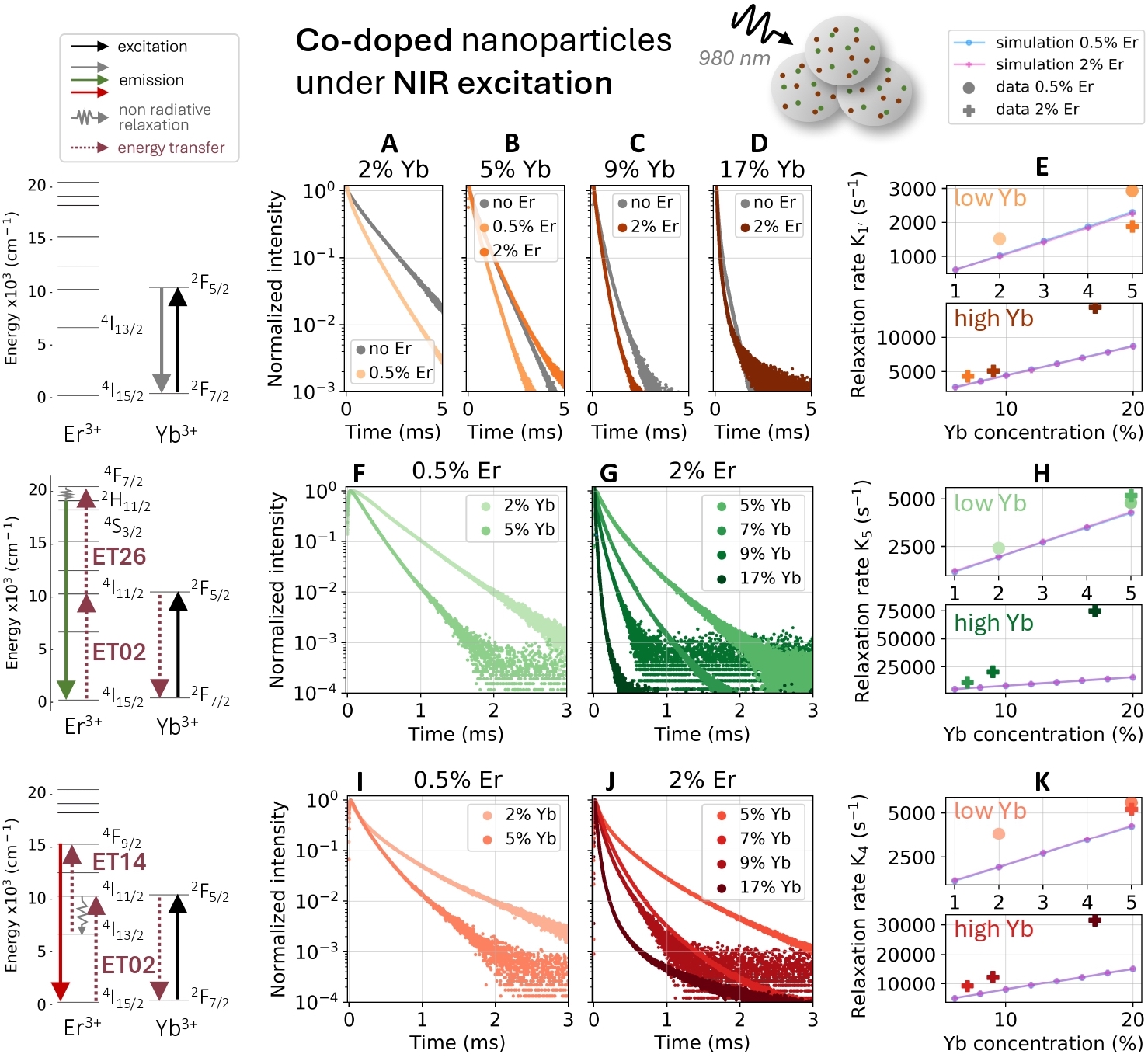}
	\caption{ \textbf{Co-doped nanoparticles - under 980 nm excitation (ETU).} (A,B,C,D) Luminescence decays of Yb$^{3+}$ excite state of nanopaticle singly-doped and co-doped with 2 \%, 5 \%, 9 \% and 17\% Yb. (E) Variation of the relaxation rate $K_{1'}$ as a function of Yb$^{3+}$ concentration. (F,G) Luminescence decays of Er$^{3+}$ green-emitting levels ($^2$H$_{11/2}$,$^4$S$_{3/2}$) through ETU excitation of nanoparticles co-doped with 0.5 \% and 2 \% Er. (H) Variation of the relaxation rate $K_5$ as a function of Yb$^{3+}$ concentration. (I,J) Luminescence decay of red-emitting level ($^4$F$_{9/2}$) through ETU excitation of nanoparticles co-doped with 0.5 \% and 2 \% Er. (K) Variation of the relaxation rate $K_4$ as a function of Yb$^{3+}$ concentration.}
	\label{fig:panel4}
\end{figure}

After characterizing interactions within each ion species (\Ybt–\Ybt\ and \Ert–\Ert) and quantifying key processes such as NR decays and CR rates, we now turn to the investigation of co-doped nanoparticles. We investigate ET from Yb$^{3+}$ to \Ert\, as well as BT from Er$^{3+}$ to \Ybt. To this end, nanoparticles doped with 0.5\% and 2\% \Ert, and varying Yb$^{3+}$ concentrations, are analyzed to assess the influence of Yb$^{3+}$ content on the excitation and relaxation dynamics. We begin by characterizing the key Er$^{3+}$ energy levels through direct excitation (Figure \ref{fig:panel3}), before examining their behavior under up-conversion excitation at 980 nm (Figure \ref{fig:panel4}). Table \ref{tab:chap5:tau_yber} summarizes experimental lifetimes of various energy levels measured in the co-doped nanoparticles. 

\begin{table}[h!]
\begin{centering}
\begin{tabular}{c|c|c|cccccc}                           \multicolumn{3}{c|}{} &      \multicolumn{6}{c}{Er concentration}                         \\                        \multicolumn{3}{c|}{}                     & 0.5 \% & 0.5 \% & 2 \%  & 2 \%  & 2 \%  & 2 \%  \\ \cline{4-9}
 \multicolumn{3}{c|}{} &      \multicolumn{6}{c}{Yb concentration}                         \\ \textit{ion} &
\textit{level} & \textit{excitation}                                                                                                       & 2 \%   & 5 \%   & 5 \%  & 7 \%  & 9 \%  & 17 \% \\ \hline
 \multirow{5}{*}{Er$^{3+}$}& \multirow{2}{*}{$^2$H$_{11/2}$,$^4$S$_{3/2}$} & direct     & 0.058  & 0.035  & 0.030 & 0.019 & 0.016 & 0.011 \\                                                                 &        & UC         & 0.41  & 0.21  & 0.19 & 0.09 & 0.05& 0.01 \\ \cline{2-9}
 &\multirow{2}{*}{$^4$F$_{9/2}$}                                           & direct     & 0.018  & 0.017  & 0.017 & 0.016 & 0.016 & 0.013 \\                                                              &           & UC         & 0.26  & 0.18  & 0.19 & 0.11 & 0.08 & 0.03 \\ \cline{2-9}             &                $^4$I$_{13/2}$                               & direct     & 7.8  & 7.4  & 6.7 & 4.4 & 6.6 & 4.3 \\ \hline   Yb$^{3+}$& $^2$F$_{5/2}$                                                         & direct     & 0.66  & 0.34  & 0.53 & 0.23 & 0.20 & 0.07
\end{tabular}
\caption{Yb$^{3+}$ and Er$^{3+}$ excited state lifetimes (ms) for different doping concentrations in co-doped nanoparticles.}
\label{tab:chap5:tau_yber}
\end{centering}
\end{table}

\textbf{Green emission - direct excitation.} Figure \ref{fig:panel3}.A (resp. \ref{fig:panel3}.B) shows luminescence decays corresponding to the $^2$H$_{11/2}$,$^4$S$_{3/2}$ $\rightarrow$ $^4$I$_{15/2}$ transition in co-doped nanoparticles containing 0.5\% Er$^{3+}$ (resp. 2\% Er)  and varying Yb$^{3+}$ concentrations.

An increase in Yb$^{3+}$ concentration leads to a slight shortening of the decay times. For instance, in nanoparticles with 17\% Yb, the lifetime $\tau$ is reduced by a factor of 4 compared to Yb-free nanoparticles. This quenching effect is attributed to the back transfer mechanism BT51 ($^4$S$_{3/2}$, $^2$F$_{7/2}$) $\rightarrow$ ($^4$I$_{13/2}$, $^2$F$_{5/2}$), with an energy gap of 1521 cm$^{-1}$. The large energy mismatch explains why concentration quenching remains relatively moderate.

In the series of nanoparticles doped with 2\% Er, a long tail appears in the luminescence decay, indicating an ETU process from the $^2$F$_{5/2}$  level once it is populated by back transfer. This suggests that $^2$F$_{5/2}$ acts as a reservoir with its long characteristic time.

Figure \ref{fig:panel2}.C illustrates the variation of the relaxation rate $K_5$ as a function of Yb$^{3+}$ concentration. For both Er$^{3+}$ concentrations (0.5\% and 2\%), $K_5$ appears to vary linearly with Yb$^{3+}$ concentration, which is consistent with the dominance of the BT mechanism. The relaxation rate can be expressed as: $K_5 = k_{\mathrm{R}50}+k_{\mathrm{R}51}+k_{\mathrm{NR}54} + k_{\mathrm{CR50}} n_{0} + k_{\mathrm{BT51}} n'_{0}$, where $k_{\mathrm{BT51}}$ represents the energy transfer rate of BT51 mechanism. A linear fit, with $k_{\mathrm{BT51}}$ as a free parameter, yields $k_{\mathrm{BT51}}=1.8\times10^{-17}$ cm$^3$.s$^{-1}$. The other parameters were fixed from the singly-doped nanoparticles section (see Table \ref{tab:chap5:tau_er_np}).

The relaxation rates $K_5$ obtained from the simulation, plotted in blue in Figure \ref{fig:panel2}.C, show good agreement with experimental data for both series of Er$^{3+}$ doping levels.

\textbf{Red emission - direct excitation.} Figure \ref{fig:panel3}.D (resp. \ref{fig:panel3}.E) shows luminescence decays corresponding to the $^4$F$_{9/2}$ $\rightarrow$ $^4$I$_{15/2}$ transition in co-doped nanoparticles containing 0.5\% Er$^{3+}$ (resp. 2\% Er)  and varying Yb$^{3+}$ concentrations.

Only a minor change is observed as the Yb$^{3+}$ concentration increases. For example, in nanoparticles with 17\% Yb, the lifetime $\tau$ is reduced by a factor of 1.5 compared to Yb-free nanoparticles. This quenching effect can attributed to the back transfer mechanism BT40 ($^4$F$_{9/2}$, $^2$F$_{7/2}$) $\rightarrow$ ($^4$I$_{15/2}$, $^2$F$_{5/2}$), which has a large energy gap of 4979 cm$^{-1}$. The substantial energy mismatch explains why concentration quenching remains minimal.

Figure \ref{fig:panel2}.F illustrates the variation of the relaxation rate $K_4$ as a function of Yb$^{3+}$ concentration. For both Er$^{3+}$ concentrations (0.5\% and 2\%), $K_4$ appears to vary linearly with Yb$^{3+}$ concentration, indicating that the BT mechanism is the dominant process. We express: $K_4 = k_{\mathrm{R}40}+k_{\mathrm{NR}43} +k_{\mathrm{BT40}} n'_{0}$, where $k_{\mathrm{BT40}}$ is the energy transfer rate of BT40 process. A linear fit with $k_{\mathrm{BT40}}$ as a free parameter, gives $k_{\mathrm{BT40}}=7.0\times10^{-18}$ cm$^3$.s$^{-1}$. 

The simulation results in relaxation rates $K_4$, plotted in blue in Figure \ref{fig:panel2}.F, that are similar to the experimental data.

The visible emission spectra of the co-doped nanoparticles under 520~nm excitation (Figure~S\ref{fig:si_spectra_520nm}) show an increasing R/G ratio with rising Yb$^{3+}$ concentration, consistent with previous reports~\cite{Vetrone2004,zhang2015observation}. As in the singly-doped samples, this trend is attributed to ion–ion interactions that more efficiently populate the $^4$F$_{9/2}$ level at higher Yb$^{3+}$ content, through pathways other than multiphonon relaxation from the $^2$H$_{11/2}$ and $^4$S$_{3/2}$ levels. In particular, the BT51 mechanism effectively populates both the $^4$I$_{13/2}$ and $^2$F$_{5/2}$ levels, enhancing the ETU process ($^4$I$_{13/2}$, $^2$F$_{5/2}$)~$\rightarrow$~($^4$F$_{9/2}$, $^2$F$_{7/2}$). Additionally, the CR51 interaction may also contribute to the observed increase in the R/G ratio. As a result of these mechanisms, the $^4$F$_{9/2}$ level becomes more populated relative to the $^4$S$_{3/2}$ level at high Yb$^{3+}$ concentrations.

\textbf{1.5 $\mu$m emission - direct excitation.} Figure \ref{fig:panel3}.G (resp. \ref{fig:panel3}.H) shows luminescence decays corresponding to the $^4$I$_{13/2}$ $\rightarrow$ $^4$I$_{15/2}$ transition in co-doped nanoparticles containing 0.5 \% Er$^{3+}$ (resp. 2\% Er) and different Yb$^{3+}$ concentrations.

Varying the Yb$^{3+}$ concentration does not induce any significant change. The decay profiles of nanoparticles doped with 0.5\% Er$^{3+}$ remain nearly identical, with a lifetime of approximately 8 ms. In the case of nanoparticles doped with 2\% Er, those containing 5\% and 9\% of Yb$^{3+}$ exhibit slightly longer lifetimes ($\approx$ 6.7 ms) compared to the others ($\approx$ 4 ms). However, this variation is not systematically correlated with Yb$^{3+}$ concentration, as shown in Figure \ref{fig:panel2}.I, which depicts the relaxation rate $K_1$ as a function of Yb$^{3+}$ content.

It is worth noting that energy level dynamics with such long lifetimes (on the order of several milliseconds) are highly sensitive to additional relaxation mechanisms, such as energy transfer to defects or impurities. Although the concentration of these defects and impurities may vary slightly between syntheses, potentially influencing the measured lifetimes, the nearly constant decay times observed as a function of \Ybt\ concentration clearly rules out large increase of defect concentration in the co-doped particles. 

The relaxation rates $K_1$ obtained from the simulation, plotted in blue in Figure \ref{fig:panel2}.I, show relatively good agreement with experimental data for both series of Er$^{3+}$ doping levels.

\vspace{0.5cm}

In summary, the luminescence decay times and relaxation rates of co-doped nanoparticles reveal distinct energy transfer mechanisms for different emission channels. While Yb$^{3+}$ doping significantly affects green and red emissions through BT mechanisms, the long-lived 1.5 µm emission remains largely unaffected by Yb$^{3+}$ concentration, evidencing a similar concentration in quenching centers in all particles. This latter conclusion will be important in the following analyses. Table~\ref{table_et_exp} summarizes the BT rates extracted from the experimental data.

\begin{table}
\begin{centering}
\begin{tabular}{c|c}
$k_{\mathrm{BT51}}$ &  $k_{\mathrm{BT40}}$   \\
 $1.8\times10^{-17}$&$7.0\times10^{-18}$ \end{tabular}
\caption{Back energy transfer rates (cm$^3$.s$^{-1}$) estimated from experimental data, and used for the modeling.}
\label{table_et_exp}
\end{centering}
\end{table}

\textbf{1 $\mu$m emission - 980 nm excitation.} Figure \ref{fig:panel4}.A, \ref{fig:panel4}.B, \ref{fig:panel4}.C, and \ref{fig:panel4}.D show luminescence decays corresponding to the $^2$F$_{5/2}$ $\rightarrow$ $^2$F$_{7/2}$ transition in co-doped nanoparticles containing 2, 5, 9, 17 \% Yb$^{3+}$  and varying Er$^{3+}$ concentrations.

The decays show that Yb$^{3+}$ emission is slightly quenched by Er$^{3+}$ which means that BT mechanisms are less efficient than Er$^{3+}$ relaxations. In this case, Er$^{3+}$ ions act as quenchers and reduces Yb$^{3+}$ lifetime. This is also confirms by the modeling showing a decreasing Yb$^{3+}$ lifetime with increasing \Ert\ concentration (Figure S\ref{fig:si_sim_yb}).

To quantitatively  evaluate Yb-Er interactions, we will consider both decay of Yb$^{3+}$ excited state and decays of the Er$^{3+}$ green and red emitting levels, all after excitation of Yb$^{3+}$ transition. This is needed because several ET and BT processes are involved in the dynamics of \Ybt\ excited state.  

\textbf{Green emission - 980 nm excitation (ETU).} Figure \ref{fig:panel4}.F (resp. \ref{fig:panel4}.G) shows luminescence decays corresponding to the $^2$H$_{11/2}$,$^4$S$_{3/2}$ $\rightarrow$ $^4$I$_{15/2}$ transition after up-conversion excitation in co-doped nanoparticles containing 0.5 \% Er$^{3+}$ (resp. 2 \% Er)  and varying Yb$^{3+}$ concentrations. The ETU mechanism in \Ert\ and \Ybt\ co-doped materials is well known and occurs via two successive energy transfers from two excited \Ybt\ ions to a nearby \Ert\ ion \cite{auzel2004}. The process is detailed in Supplementary Information.

As expected, it can be observed that the luminescence decay times of the different nanoparticles are longer in this case than when measured with direct excitation (see Figures \ref{fig:panel3}.A and \ref{fig:panel3}.B). For instance, nanoparticles doped with 2\% Yb$^{3+}$ and 0.5\% Er$^{3+}$ exhibit a decay time of 0.41 ms with ETU excitation, compared to 0.058 ms under direct excitation. This difference is attributed to the millisecond-range lifetimes of the Yb$^{3+}$ excited state and the $^4$I$_{11/2}$ level of Er, which serve as energy reservoirs in the process. This further confirms that the levels are populated via ETU rather than excited-state absorption (ESA).

Additionally, the decay times of the $^2$H$_{11/2}$ and $^4$S$_{3/2}$ levels decrease with increasing Yb$^{3+}$ concentration, as previously observed in ceramics \cite{kimura2021}, and nanocrystals \cite{Vetrone2004}. This trend is mainly attributed to the reduced lifetime of the $^2$F$_{5/2}$ level, which is already observed in singly-doped nanoparticles. However, additional mechanisms present in co-doped nanoparticles, such as ET and BT, also influence the overall dynamics. This is confirmed by the modeling indicating that increasing the Er$^{3+}$ concentration leads to higher relaxation rates of the green-emitting levels through ETU excitation (Figure S\ref{fig:si_sim_green}).

Optimizing UC luminescence requires balancing two competing effects: increasing rare-earth ion concentrations enhances absorption and energy transfer probability, but excessively high concentrations lead to concentration quenching, as observed in our study of singly-doped nanoparticles. Based on our results, a composition around 2\% \Ybt\ and 0.5\% \Ert\ likely provides an optimal trade-off for maximizing ETU efficiency. This hypothesis could be verified by quantitatively comparing the emission intensities across the different doping concentrations.

\textbf{Red emission - 980 nm excitation (ETU).} Figure \ref{fig:panel4}.I (resp. \ref{fig:panel4}.J) shows luminescence decays corresponding to the $^4$F$_{9/2}$ $\rightarrow$ $^4$I$_{15/2}$ transition after NIR excitation in co-doped nanoparticles containing 0.5\% Er$^{3+}$ (resp. 2\% Er)  and varying Yb$^{3+}$ concentrations.

As observed for the green-emitting levels, the $^4$F$_{9/2}$ level exhibits longer decay times through ETU excitation compared to direct excitation. For instance, nanoparticles doped with 2\% Yb$^{3+}$ and 0.5\% Er$^{3+}$ display a lifetime of approximately 0.26 ms when excited at 976 nm, whereas direct excitation at 655 nm results in a significantly shorter lifetime of 0.018 ms. These relatively long lifetimes further confirm that the red-emitting level is predominantly populated via energy transfer from longer-lived excited states.

Similar to the behavior of the $^2$H$_{11/2}$,$^4$S$_{3/2}$ levels, the fluorescence decay time of the $^4$F$_{9/2}$ level decreases with increasing Yb$^{3+}$ concentration. This trend is primarily attributed to the reduction in the Yb$^{3+}$ $^2$F$_{5/2}$ lifetime but also to other \Ybt-\Ert\ interactions since modeling shows that the red-emitting level lifetime decreases with increasing \Ert\ concentration (Figure S\ref{fig:si_sim_red}).

\textbf{ETU  modeling.} To accurately model Yb–Er interactions following ETU excitation, we now focus on the key interactions: ET02, ET26, ET14, and BT20. Figure~\ref{fig:panel4}.E, \ref{fig:panel4}.H and \ref{fig:panel4}.K show the relaxation rates $K_2$, $K_5$, and $K_4$ as a function of Yb concentration, obtained from simulations using the following parameters: $k_{\mathrm{ET02}} = k_{\mathrm{ET26}} = 10^{-16}$~cm$^3\cdot$s$^{-1}$, $k_{\mathrm{ET14}} = 10^{-19}$~cm$^3\cdot$s$^{-1}$, and $k_{\mathrm{BT20}} = 10^{-15}$~cm$^3\cdot$s$^{-1}$.

These ET and BT rates do not results from a fit, but were varied within a reasonable range (from $10^{-14}$ to $10^{-19}$~cm$^3\cdot$s$^{-1}$), and no combination yielded significantly better agreement between the simulated and experimental data than those shown in Figure~\ref{fig:panel4}.E, ~\ref{fig:panel4}.H, and ~\ref{fig:panel4}.K. These parameters were found to be interdependent and also sensitive to the initial condition.

Comparing the simulation results with the experimental data, the model accurately reproduces the behavior observed at  low to medium Yb$^{3+}$ concentrations (<5\%) for both 0.5\% and 2\% Er. We thus obtain a comprehensive and accurate description of Er-Yb interactions in which energy transfer rates are mostly individually determined from specific measurements. In contrast with modeling were all parameters are simultaneously optimized, it can be expected that our approach leads to more reliable and physically meaningful parameters and in turn predictions of decay dynamics for a broad range of concentrations with fast computations. Interestingly, the fact that rate equations are able to reproduce experimental decay rates indicates that the microscopic variations in ions' relative positions can be taken into account to a large extent by averaged values. 

At very high Yb$^{3+}$ concentrations, it can be noted that the experimental relaxation rates of all the three levels studied here, $^4$S$_{3/2}$, $^4$F$_{9/2}$, and $^2$F$_{5/2}$, are significantly faster than those predicted by the model (Figure \ref{fig:panel4}, right column, lower graphs). As in the analysis of the singly-doped particles, this is attributed to the onset of fast energy migration between \Ybt\ ions that effectively increases Er-Yb energy transfer rates. Thanks to our comprehensive study, a large increase of the concentration in quenching defects at high \Ybt\ can be safely dismissed given the nearly constant \qIt\ level decays observed in all co-doped particles.

\subsection{Conclusion}

In conclusion, we have investigated energy transfer in RE doped nanoparticles by studying the dynamics of multiple transitions, doping concentrations, and excitation paths, in order to achieve a broad understanding of the phenomena involved. Specifically, we studied  \YO\ nanoparticles of 100 nm diameter singly- or  co-doped with \Ert\ and \Ybt\ in a broad range of concentrations extending from 0.5 to 17 at. \%. First, PL decays were recorded for \Ert\ green (\qSt,\dHo $\to$ \qIq), red (\qFn$\to$ \qIq) and IR (\qIo $\to$ \qIq, and \qIt $\to$ \qIq) transitions under direct excitation in the respective levels. The same experiment were performed on \Ybt\ doped nanoparticles (\dFc$\to$ \dFs\ transition). In a second series of experiments, transitions were studied in co-doped nanoparticles under direct excitation and then using ETU, a process of high interest for various applications. 

In systems such as the one we studied, a large number of interactions between RE ions and defects can contribute to decay dynamics. To separate them and extract meaningful and usable values, we analyzed our decays using rate equations models and focused on reproducing the decay rates defined as $1/\tau$ where 2$\tau$ is the time for which the initial intensity has been reduced by $1/e^2=0.14$. This was chosen as a balance between short and long-time behaviors for non-exponential decays. 

This modeling allowed us to reproduce direct excitation data and determine energy transfer relaxation rates in singly-doped nanoparticles in a broad range of concentrations. Energy transfers between RE ions but also towards quenching centers were taken into account, the latter being promoted by energy migration and thus important at high RE concentration. These relaxation rates were then used in the modeling  of decays in co-doped particles, enabling extracting \Ert-\Ybt specific interactions and reproducing well experimental decays rates under direct excitation. For ETU, simulations are in good agreement with experiments up to \Ybt\ concentrations of 5\%. At levels above 10\%, deviations were observed, as the experimental decays were clearly shorter than predicted, and are attributed to fast energy migration between \Ybt\ ions which increases Yb-Er energy transfer rates. Based on our analysis, a composition near 2\% \Ybt\ and 0.5\% \Ert\ appears optimal for maximizing ETU efficiency.

This work provides an exhaustive experimental study of interactions in a RE doped material and a modeling able to accurately predict decay rates of multiple transitions, in broad concentration ranges, and for different excitation paths. This approach opens the way to efficient optimization of PL in \Ert,
\Ybt:\YO\ nanoparticles and should be straightforward to extend to particles of different sizes, thin films, or transparent ceramics. More generally, it could be useful to analyze, design, and enhance RE doped materials in which emissions and excitations involve complex interactions.

\begin{acknowledgement}

This project has received funding from the European Research Council (ERC) under the European Union’s Horizon 2020 research and innovation programme (grant agreement No 101019234, RareDiamond) the 80 prime CNRS project Mathyq, and the USP-COFECUB programme (project Uc Ph-C 183/20).	 

\end{acknowledgement}


\newpage

\begin{suppinfo}

\subsection{Synthesis of rare-earth doped Y$_2$O$_3$ nanoparticles}

In a typical synthesis \cite{deoliveiralimaInfluenceDefectsSubA2015}, the required amounts of yttrium nitrate, erbium or ytterbium nitrate, and urea were dissolved in deionized water to a total volume of 252 mL, under vigorous magnetic stirring for 15 minutes at room temperature, within a Teflon reactor. The sealed reactor was then subjected to hydrothermal treatment in a drying oven at 85 °C for 24 hours.

Following the reaction, the resulting precipitate was collected by centrifugation at 13,000 rpm for 10 minutes at 4 °C, using multiple washing steps with deionized water to remove the byproducts. Anhydrous ethanol was used for the final washing of the nanoparticles. The precursor was suspended in anhydrous ethanol and subsequently dried at 80 °C for 24 hours to obtain Y(OH)CO$_3$.nH$_2$O. Crystalline Y$_2$O$_3$:RE$^{3+}$ nanoparticles (RE$^{3+}$ = Er$^{3+}$ or Yb$^{3+}$) were formed via two-step annealing, initially at 800 °C for 6 hours, followed by a second treatment at 1200 °C for 18 hours using a heating rate of 3 ºC. min$^{-1}$.

\subsection{Simulation}

\subsubsection{Radiative and non-radiative relaxation rates}

Radiative rates from each excited state level and branching ratio are taken from literature \cite{weber1968,sardar2007}. Non-radiative decay rates are estimated from low-concentration experiments or fitted experimental data. The parameters are displayed in Table \ref{tab:chap5:sim_rate_r_nr}.

\begin{table}
\begin{centering}
\begin{tabular}{c|c|c|c|c|c}
$k_{\mathrm{R}51}$& $k_{\mathrm{R}50}$& $k_{\mathrm{R}40}$& $k_{\mathrm{R}30}$& $k_{\mathrm{R}20}$& $k_{10}$        \\
$4.2\times10^2$ & $9.8\times10^2$ & $1.7\times10^3$ & $2.2\times10^2$ & $1.5\times10^2$ & 186\\ \hline
$k_{\mathrm{NR}65}$& $k_{\mathrm{NR}54}$& $k_{\mathrm{NR}43}$& $k_{\mathrm{NR}32}$& $k_{\mathrm{NR}21}$&    $k_{1'0'}$             \\
$1\times10^7$ & $5.5\times10^3$& $5.5\times10^4$& $3.8\times10^5$& 448&   173\end{tabular}
\caption{Radiative and non-radiative relaxation rates (s$^{-1}$) of Er$^{3+}$ and Yb$^{3+}$ ions used for the modeling.}
\label{tab:chap5:sim_rate_r_nr}
\end{centering}
\end{table}

\subsubsection{Energy transfer rates}

Some of the energy transfer rates such as $k_{\mathrm{ET1'0'd}}$, $k_{\mathrm{CR50}}$, $k_{\mathrm{5HC}}$, $k_{\mathrm{2HC}}$, $k_{\mathrm{1HC}}$, $k_{\mathrm{BT51}}$ and $k_{\mathrm{BT40}}$ have been determined directly from experimental data and are reminded in Table \ref{table_si_et_exp}.

\begin{table}
\begin{centering}
\begin{tabular}{c|c|c|c|c|c|c}
$k_{\mathrm{ET1'0'd}}$& $k_{\mathrm{CR50}}$& $k_{\mathrm{BT51}}$ &  $k_{\mathrm{BT40}}$ & $k_{\mathrm{5HC}}$& $k_{\mathrm{2HC}}$& $k_{\mathrm{1HC}}$    \\
$2.2\times10^{-18}$ & $4.9\times10^{-17}$ & $1.8\times10^{-17}$&$7.0\times10^{-18}$& $1.6\times10^{-16}$ & $9.4\times10^{-19}$ & $3.1\times10^{-18}$ \end{tabular}
\caption{Energy transfer rates (cm$^3$.s$^{-1}$) estimated from experimental data used for the modeling.}
\label{table_si_et_exp}
\end{centering}
\end{table}

Key energy transfer rates for UC mechanism such as $k_{\mathrm{ET02}}$,  $k_{\mathrm{ET26}}$, $k_{\mathrm{ET14}}$, $k_{\mathrm{BT20}}$ have been fixed by comparing experimental data with simulation results and are reported in Table \ref{table_et_sim}.

\begin{table}
\begin{centering}
\begin{tabular}{c|c|c|c}
$k_{\mathrm{ET02}}$& $k_{\mathrm{ET26}}$& $k_{\mathrm{ET14}}$& $k_{\mathrm{BT20}}$    \\
$10^{-16}$ & $10^{-16}$ & $10^{-19}$& $10^{-15}$\end{tabular}
\caption{Energy transfer rates (cm$^3$.s$^{-1}$) estimated from experimental data used for the modeling.}
\label{table_et_sim}
\end{centering}
\end{table}

A wide variety of other ion–ion interactions can occur, due to the large number of available energy levels in Er$^{3+}$ ions. To simplify simulations and improve the interpretability of the mechanisms involved, it is useful to limit the model to the dominant interactions. This selection can be guided by the energy gaps associated with the transfers, as well as by previously conducted experimental analyses.

The probability of an energy transfer involving a mismatch $\Delta E$ decreases exponentially as $\exp(-\beta \Delta E)$, where $\beta$ is a material-dependent parameter. For Y$_2$O$_3$ matrices, $\beta$ has been estimated to be $2.5 \times 10^{-3}~\mathrm{cm}$ \cite{yamada1972}. In that study, conducted at 77~K, the probability of resonant energy transfer was evaluated to be approximately $10^5~\mathrm{s^{-1}}$ for a donor concentration of 4\%, corresponding to about $7 \times 10^{20}~\mathrm{cm^{-3}}$. This implies a resonant energy transfer coefficient of $k_{\mathrm{ET}} \approx 1.4 \times 10^{-16}~\mathrm{cm^3 \cdot s^{-1}}$.

Although this approximation does not explicitly account for the variation in coupling strengths between different transitions, it offers a reasonable estimate in most cases \cite{yamada1972}.

The remaining energy transfer rates, which could not be determined experimentally, are estimated using this method and are reported in Table \ref{tab:chap5:energytransfer_energygap}.

\begin{table}[h]
\begin{centering}
\begin{tabular}{cccccc}
                                                                                                                         & transition 1                                & transition 2                                & \begin{tabular}[c]{@{}c@{}}$\Delta E$ \\  (cm$^{-1}$)\end{tabular}  &  notation     & \begin{tabular}[c]{@{}c@{}}rate\\  (cm$^{3}$.s$^{-1}$)\end{tabular}     \\ \hline 

\multirow{2}{*}{\begin{tabular}[c]{@{}c@{}}back energy  transfer\\[5pt] Er$^{3+}$ $\rightarrow$ Yb$^{3+}$\end{tabular}} & $^4$F$_{7/2}$ $\rightarrow$ $^4$I$_{11/2}$  & $^2$F$_{7/2}$ $\rightarrow$ $^2$F$_{5/2}$   &               100         & $k_{\mathrm{BT62}}$ & $1.4\times10^{-16}$\\[5pt]
                                                                                                                       & $^4$F$_{9/2}$ $\rightarrow$ $^4$I$_{13/2}$  & $^2$F$_{7/2}$ $\rightarrow$ $^2$F$_{5/2}$   &            -1479            &  &0  \\[5pt]
                                    \hline
\multirow{8}{*}{\begin{tabular}[c]{@{}c@{}}cross-relaxation\\ Er$^{3+}$ $\rightarrow$ Er$^{3+}$\end{tabular}}          & $^4$F$_{7/2}$ $\rightarrow$ $^4$I$_{11/2}$  & $^4$I$_{15/2}$ $\rightarrow$ $^4$I$_{11/2}$ &           119             &  &0  \\[5pt]
                                                                                                                         & $^4$I$_{11/2}$ $\rightarrow$ $^4$I$_{15/2}$ & $^4$I$_{11/2}$ $\rightarrow$ $^4$F$_{7/2}$  &          -119              &  &0   \\[5pt]
                                                                                                                         & $^4$F$_{7/2}$ $\rightarrow$ $^4$F$_{9/2}$   & $^4$I$_{11/2}$ $\rightarrow$ $^4$F$_{9/2}$  &              196          & $k_{\mathrm{CR}62}$& $1.4\times10^{-16}$ \\[5pt]
                                                                                                                    & $^2$H$_{11/2}$ $\rightarrow$ $^4$F$_{9/2}$  & $^4$I$_{13/2}$ $\rightarrow$ $^4$I$_{11/2}$ &             245           & $k_{\mathrm{CR}51}$ & $1.4\times10^{-16}$\\[5pt]
                                                                                                                         & $^4$I$_{11/2}$ $\rightarrow$ $^4$I$_{13/2}$ & $^4$F$_{9/2}$ $\rightarrow$ $^4$S$_{3/2}$   &             615           & $k_{\mathrm{CR}42}$& $3.0\times10^{-17}$ \\[5pt]
                                                                                                                  & $^4$I$_{13/2}$ $\rightarrow$ $^4$I$_{15/2}$ & $^4$I$_{13/2}$ $\rightarrow$ $^4$I$_{9/2}$  &              628          &  $k_{\mathrm{CR}11}$&$2.9\times10^{-17}$  \\[5pt] 
\end{tabular}
\caption{Energy transfer rates estimated based on the energy gap.}
\label{tab:chap5:energytransfer_energygap}
\end{centering}
\end{table}

The cross-relaxation process ($^4$I$_{11/2}$, $^4$I$_{11/2}$) $\rightarrow$ ($^4$I$_{15/2}$, $^4$F$_{7/2}$), as well as its reverse process ($^4$F$_{7/2}$, $^4$I$_{15/2}$) $\rightarrow$ ($^4$I$_{11/2}$, $^4$I$_{11/2}$), can be considered negligible. These mechanisms are analogous to the direct energy transfer ($^2$F$_{5/2}$, $^4$I$_{11/2}$) $\rightarrow$ ($^2$F$_{7/2}$, $^4$F$_{7/2}$) and the back energy transfer ($^4$F$_{7/2}$, $^2$F$_{7/2}$) $\rightarrow$ ($^4$I$_{11/2}$, $^2$F$_{5/2}$) involving Yb$^{3+}$ ions, but are less likely to occur since the $^4$I$_{15/2}$ $\rightarrow$ $^4$I$_{11/2}$ transition is approximately seven times weaker than the $^2$F$_{7/2}$ $\rightarrow$ $^2$F$_{5/2}$ transition~\cite{weber1968}.

\subsubsection{Rate equations}

The population rate equations used in the modeling are the following \cite{auzel1973}:
\setlength{\jot}{0pt}
\newpage
 {\footnotesize
\begin{flalign}
    \frac{dn_{1'}}{dt}=&
    -k_{1'0'}n_{1'}
    -k_{\mathrm{ET02}}n_{0}n_{1'}
    -k_{\mathrm{ET26}}n_{2}n_{1'}
    -k_{\mathrm{ET14}}n_{1}n_{1'}
    \\ \nonumber
    &+k_{\mathrm{BT20}}n_{2}n_{0'}
    +k_{\mathrm{BT62}}n_{6}n_{0'}
    +k_{\mathrm{BT51}}n_{5}n_{0'} 
    +k_{\mathrm{BT40}}n_{4}n_{0'}\\\nonumber
   & -k_{\mathrm{ET1'0'd}}n_{0'}n_{1'}\\\nonumber\\
    \frac{dn_{0'}}{dt}=&
    +k_{1'0'}n_{1'}
    +k_{\mathrm{ET02}}n_{0}n_{1'}
    +k_{\mathrm{ET26}}n_{2}n_{1'}
    +k_{\mathrm{ET14}}n_{1}n_{1'}
    \\ \nonumber
    &-k_{\mathrm{BT20}}n_{2}n_{0'}
    -k_{\mathrm{BT62}}n_{6}n_{0'}
    -k_{\mathrm{BT51}}n_{5}n_{0'} 
    -k_{\mathrm{BT40}}n_{4}n_{0'}\\\nonumber
   &+k_{\mathrm{ET1'0'd}}n_{0'}n_{1'}\\ \nonumber \\
        \frac{dn_{6}}{dt}=&
    -k_\mathrm{NR65}n_{6}
    +k_{\mathrm{ET26}}n_{2}n_{1'}
    -k_{\mathrm{BT62}}n_{0'}n_{6}
    -k_{\mathrm{CR62}}n_{2}n_{6}
    \\ \nonumber \\
            \frac{dn_{5}}{dt}=&
    +k_\mathit{NR65}n_{6}
    -k_{\mathrm{R51}}n_{5}
    -k_{\mathrm{R50}}n_{5}
    -k_\mathit{NR54}n_{5} \\ \nonumber
    &-k_{\mathrm{BT51}}n_{0'}n_{5}
    -k_{\mathrm{CR51}}n_{1}n_{5}
    -k_{\mathrm{CR50}}n_{0}n_{5}
    +k_{\mathrm{CR42}}n_{4}n_{2}\\ \nonumber
    &(-k_{\mathrm{5HC}}n_{0}n_{5})
    \\ \nonumber \\
    \frac{dn_{4}}{dt}=&
    +k_\mathit{NR54}n_{5}
    -k_{\mathrm{R40}}n_{4}
    -k_\mathit{NR43}n_{4} 
    +k_{\mathrm{ET14}}n_{1}n_{1'}\\ \nonumber
    &+2k_{\mathrm{CR62}}n_{6}n_{2}
    +k_{\mathrm{CR51}}n_{1}n_{5}
    -k_{\mathrm{CR42}}n_{2}n_{4}
    -k_{\mathrm{BT40}}n_{0'}n_{4}
    \\ \nonumber \\
        \frac{dn_{3}}{dt}=&
    +k_\mathit{NR43}n_{4}
    -k_{\mathrm{R30}}n_{3}
    -k_\mathit{NR32}n_{3} 
    +k_{\mathrm{CR50}}n_{5}n_{0}
    +k_{\mathrm{CR11}}n_{1}n_{1}
     \\ \nonumber \\
    \frac{dn_{2}}{dt}=&
    +k_\mathit{NR32}n_{3}
    -k_{\mathrm{R20}}n_{2}
    -k_\mathit{NR21}n_{2} 
    +k_{\mathrm{ET02}}n_{0}n_{1'}\\ \nonumber
    &-k_{\mathrm{ET26}}n_{1'}n_{2}
    +k_{\mathrm{BT62}}n_{0'}n_{6}
    -k_{\mathrm{BT20}}n_{0'}n_{2}
    -k_{\mathrm{CR62}}n_{6}n_{2}
    \\ \nonumber
    &-k_{\mathrm{CR42}}n_{4}n_{2}
    +k_{\mathrm{CR51}}n_{5}n_{1}
    (-k_{\mathrm{2HC}}n_{0}n_{2})
        \\ \nonumber \\
    \frac{dn_{1}}{dt}=&
    +k_\mathit{NR21}n_{2}
    +k_{\mathrm{R51}}n_{5}
    -k_{10}n_{1}
    -k_{\mathrm{ET14}}n_{1'}n_{1}
    \\ \nonumber
    &+k_{\mathrm{BT51}}n_{0'}n_{5}
    -k_{\mathrm{CR51}}n_{5}n_{1} 
    +k_{\mathrm{CR50}}n_{5}n_{0}
    \\ \nonumber
    &+k_{\mathrm{CR42}}n_{4}n_{2}
    -2k_{\mathrm{CR11}}n_{1}n_{1}
   ( - k_{\mathrm{1HC}}n_{0}n_{1})
    \\ \nonumber \\
        \frac{dn_{0}}{dt}=&
    +k_{\mathrm{R50}}n_{5}
    +k_{\mathrm{R40}}n_{4}
    +k_{\mathrm{R30}}n_{3}
    +k_{\mathrm{R20}}n_{2}
    +k_{10}n_{1} \\ \nonumber
    &
    -k_{\mathrm{ET02}}n_{1'}n_{0}
    +k_{\mathrm{BT20}}n_{0'}n_{2}
    +k_{\mathrm{BT40}}n_{0'}n_{4}
-k_{\mathrm{CR50}}n_{5}n_{0}\\ \nonumber
    &
    +k_{\mathrm{CR11}}n_{1}n_{1}
    (+k_{\mathrm{5HC}}n_{0}n_{5}
    +k_{\mathrm{2HC}}n_{0}n_{2}
    +k_{\mathrm{1HC}}n_{0}n_{1})
\end{flalign}
}

with the following conditions : $\sum_{i=0}^{n-1}n_i=N_{Er}$ et $n_{0'}+n_{1'}=N_{Yb}$ and the initial conditions. We consider at $t=0$ that the excited state is populated with 1\% of the ions.

The energy transfer rates $k_{\mathrm{5HC}}$, $k_{\mathrm{2HC}}$, and $k_{\mathrm{1HC}}$ are used exclusively in the simulation of nanoparticles with Er$^{3+}$ doping concentrations > 5\%.

\subsection{Yb excited state decays fitted with Inokuti model}

Inokuti and Hirayama modeled the energy transfer dynamics in the case of dipole–dipole interactions between donors (ions excited at $t=0$) and acceptors (ions or defects not excited at $t=0$ that receive the transferred energy) using the following expression \cite{Inokuti1965}:
\begin{equation}
I(t) = I_0\ \mathrm{exp}\left(-\frac{4}{3}\pi^{3/2}N_A\sqrt{C t}\right)\mathrm{exp}(-t/\tau)
\label{f:chap5:inokuti}
\end{equation}
where $N_A$ is the acceptor concentration, $C$ represents the strength of the donor–acceptor interaction, and $\tau$ accounts for both the intrinsic donor lifetime $\tau_0$ and the energy migration lifetime $\tau_D$ due to donor–donor energy transfer before reaching an acceptor: $1/\tau = 1/\tau_0 + 1/\tau_D$. The first exponential term, which varies with $\sqrt{t}$, leads to a rapid drop in luminescence at short times, while the second term results in a mono-exponential decay at longer times.

To account for the distribution of donor environments in Yb$^{3+}$-doped systems, luminescence decays were fitted using Equation~\ref{f:chap5:inokuti}. The decay time $\tau$ was first estimated by fitting the tail of the decay curve with a mono-exponential function. Then, the parameter $N_A\sqrt{C}$ was fitted to capture the deviation from exponential behavior at short times.

\begin{figure}[h]\centering
	\includegraphics[width=0.65\textwidth]{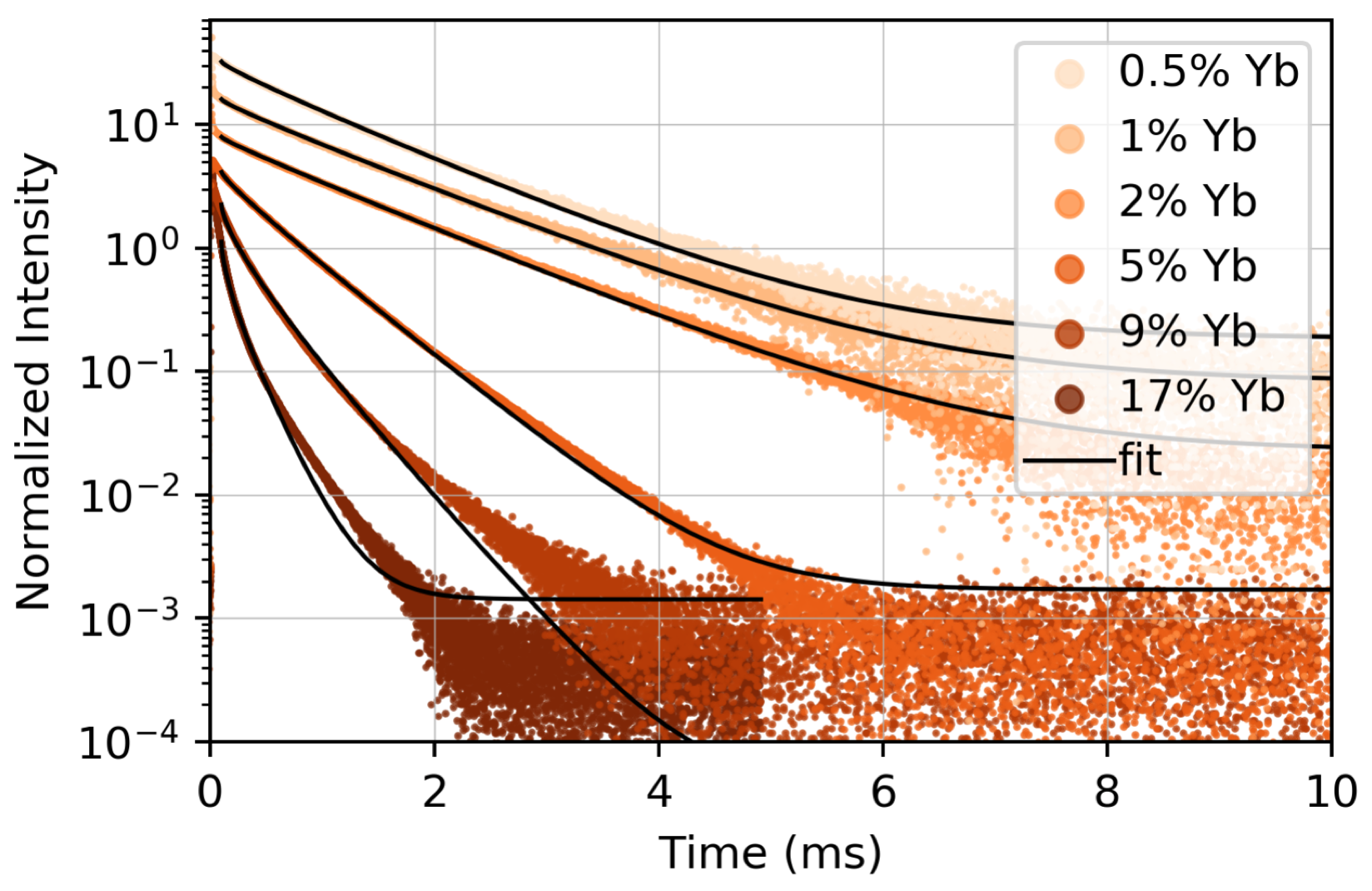}
	\caption{Luminescence decays of the \Ybt\ excited state level for various doping levels, fitted using the model developed by Inokuti and Hirayama. The decays are vertically shifted for clarity.}
	\label{fig:si_yb_inokuti}
\end{figure}

Table~\ref{table_si_inokuti} summarizes the fitting parameters used in the analysis.

\begin{table}[h!]
\begin{centering}
\begin{tabular}{c|cccccc}
                           & 0.5\% Yb             & 1\% Yb               & 2\% Yb               & 5\% Yb               & 9\% Yb               & 17\% Yb              \\ \hline
$\tau$ (ms)                & 1.24                 & 1.32                 & 1.25                 & 0.655                & 0.538                & 0.347                \\
$N_A\sqrt{C}$ (s$^{-1/2}$) & 9.6$\times$10$^{-4}$ & 7.9$\times$10$^{-4}$ & 6.2$\times$10$^{-4}$ & 1.6$\times$10$^{-3}$ & 6.0$\times$10$^{-3}$ & 1.1$\times$10$^{-2}$
\end{tabular}
\caption{Parameters used to fit the luminescence decays in Figure~\ref{fig:si_yb_inokuti}.}
\label{table_si_inokuti}
\end{centering}
\end{table}

As shown in Figure~\ref{fig:si_yb_inokuti}, the Inokuti–Hirayama model provides a reasonably good fit to the experimental decay curves across the range of doping levels. More advanced models, such as the one proposed by Yokota and Tanimoto~\cite{yokota1967}, were not used, as they introduce additional free parameters without significantly improving the quality of the fit.

\subsection{Visible emission spectra of co-doped nanoparticles under 520 nm excitation}

Figure~\ref{fig:si_spectra_520nm} shows the visible emission spectra of co-doped nanoparticles under 520~nm excitation. The emission bands are similar to those observed in Er$^{3+}$-only doped nanoparticles, with peaks around 550~nm—corresponding to the $^2$H$_{11/2}$, $^4$S$_{3/2}$~$\rightarrow$~$^4$I$_{15/2}$ transitions—and around 660~nm, associated with the $^4$F$_{9/2}$~$\rightarrow$~$^4$I$_{15/2}$ transition (see Figure~\ref{fig:panel1}.E). A clear increase in the R/G ratio is observed with increasing Yb$^{3+}$ concentration, indicating a strong influence of Yb$^{3+}$ on the population dynamics of the $^2$H$_{11/2}$, $^4$S$_{3/2}$, and $^4$F$_{9/2}$ levels. This effect is likely driven by specific interactions such as BT51 and CR51 (see Figure~\ref{fig:panel3}).

\begin{figure}[h]\centering
	\includegraphics[width=0.65\textwidth]{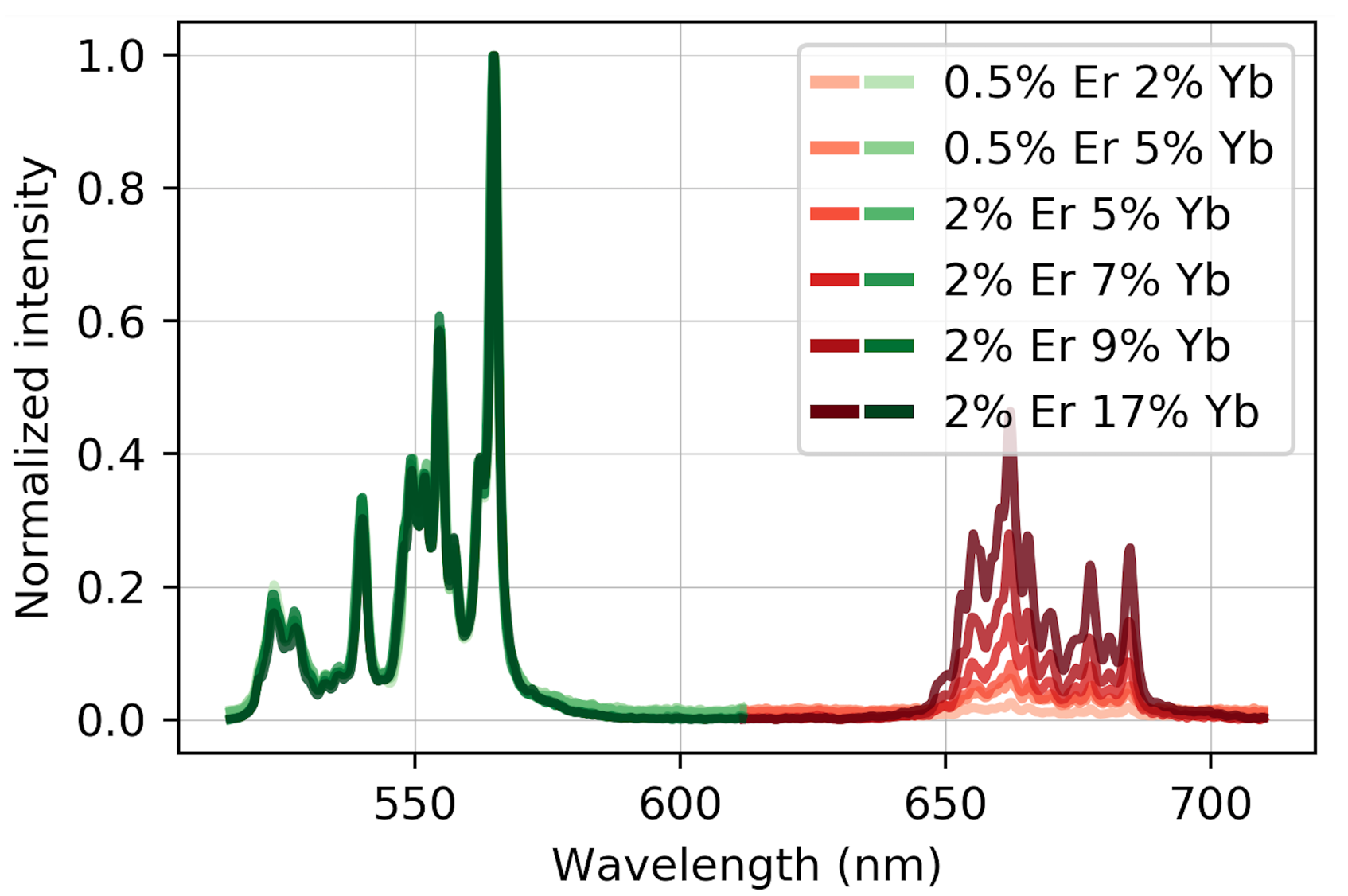}
	\caption{Emission spectra of the co-doped nanoparticles in the visible range following excitation at 520 nm.}
	\label{fig:si_spectra_520nm}
\end{figure}

\subsection{Visible emission spectra of co-doped nanoparticles under 980 nm excitation}

Figure~\ref{fig:si_spectra_980nm} shows the visible emission spectra of co-doped nanoparticles under 980~nm excitation. Emission bands are observed around 550~nm, corresponding to the $^2$H$_{11/2}$, $^4$S$_{3/2} \rightarrow {}^4$I$_{15/2}$ transitions, and around 660~nm, corresponding to the $^4$F$_{9/2} \rightarrow {}^4$I$_{15/2}$ transition. These features confirm the occurrence of UC process between Yb$^{3+}$ and Er$^{3+}$ ions.

Under near-infrared excitation, green-emitting levels of \Ert\ ions are populated via ETU from \Ybt\ ions \cite{auzel2004}. This process involves two sequential energy transfers: first, $(^2\mathrm{F}^{\mathrm{Yb}}_{5/2}, ^4\mathrm{I}^{\mathrm{Er}}_{15/2}) \rightarrow (^2\mathrm{F}^{\mathrm{Yb}}_{7/2}, ^4\mathrm{I}^{\mathrm{Er}}_{11/2})$, then $(^2\mathrm{F}^{\mathrm{Yb}}_{5/2}, ^4\mathrm{I}^{\mathrm{Er}}_{11/2}) \rightarrow (^2\mathrm{F}^{\mathrm{Yb}}_{7/2}, ^4\mathrm{F}^{\mathrm{Er}}_{7/2})$, followed by non-radiative relaxation to the green-emitting levels ($^2$H$_{11/2}$, $^4$S$_{3/2}$).
In parallel, the $^4$I$_{13/2}$ level can be populated via non-radiative decay from $^4$I$_{11/2}$. Then, another ET $(^2\mathrm{F}^{\mathrm{Yb}}_{5/2}, ^4\mathrm{I}^{\mathrm{Er}}_{13/2}) \rightarrow (^2\mathrm{F}^{\mathrm{Yb}}_{7/2}, ^4\mathrm{F}^{\mathrm{Er}}_{9/2})$ populates the red-emitting level $^4$F$_{9/2}$. In \YO\ materials, this pathway is the dominant mechanism for red emission. In contrast, for other host crystals, the red-emitting level is populated by non-radiative relaxation from the green-emitting levels, themselves populated via ETU as previously described \cite{Wu2018}.

Compared to direct excitation of the $^2$H$_{11/2}$ and $^4$S$_{3/2}$ levels—where the green emission dominates over red emission (see Figure~\ref{fig:si_spectra_520nm})—the reversed intensity ratio observed here confirms that the $^4$F$_{9/2}$ level is not mainly populated by energy transfer from the $^4$I$_{13/2}$ level.

An increase of the R/G ratio with increasing \Ybt\ concentration can be attributed to the BT51 mechanism that depopulates efficiently the green-emitting levels at high \Ybt\ concentration.

\begin{figure}[h]\centering
	\includegraphics[width=0.65\textwidth]{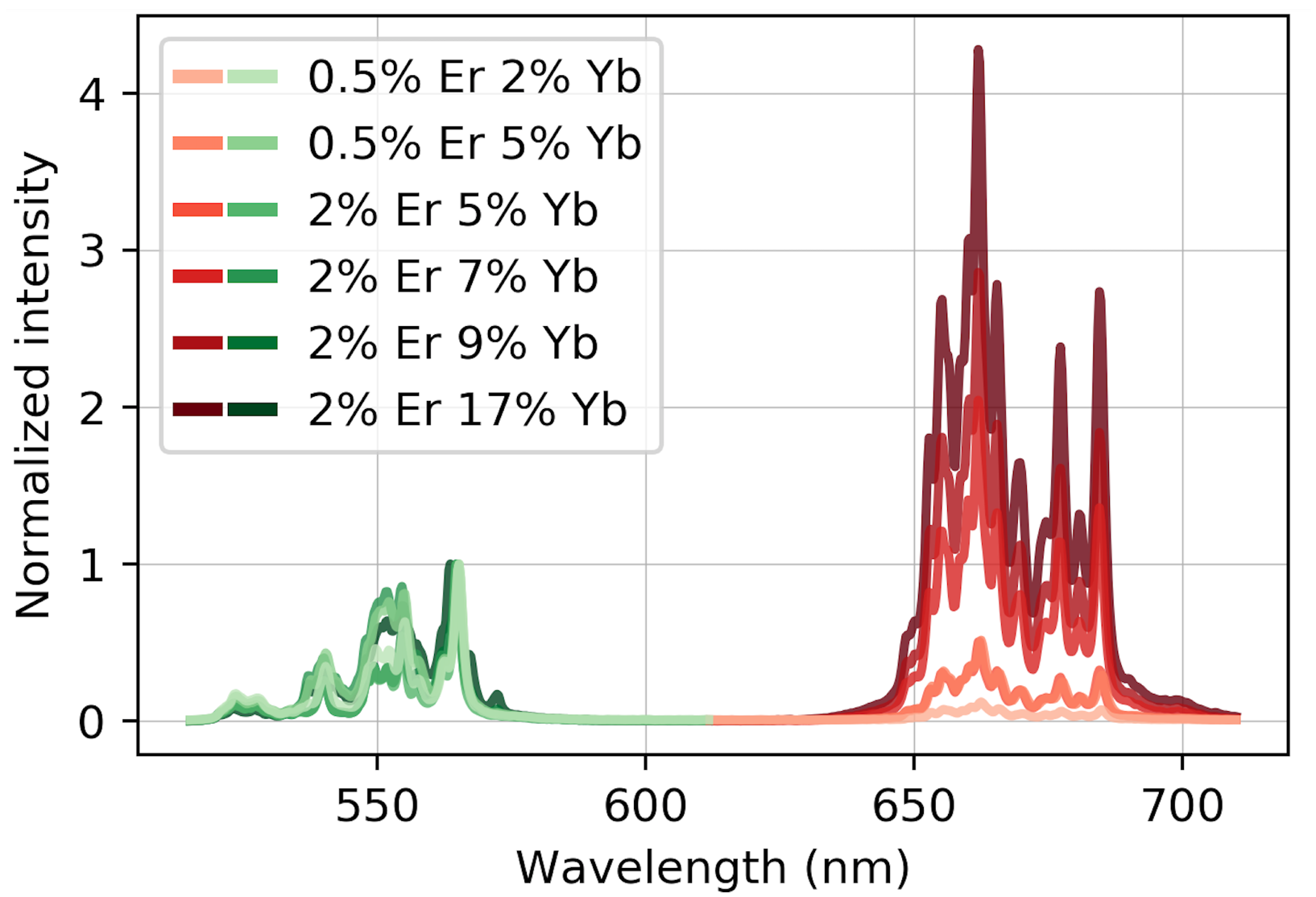}
	\caption{Emission spectra of the co-doped nanoparticles in the visible range following excitation at 980 nm.}
	\label{fig:si_spectra_980nm}
\end{figure}

\subsection{Modeling of green luminescence decay following up-conversion excitation}

Figure~\ref{fig:si_sim_uc} presents the fluorescence decays of the green-emitting levels through ETU excitation in co-doped nanoparticles. The simulated decay curves (dashed blue lines) show good agreement with the experimental data.

\begin{figure}[h]\centering
	\includegraphics[width=0.35\textwidth]{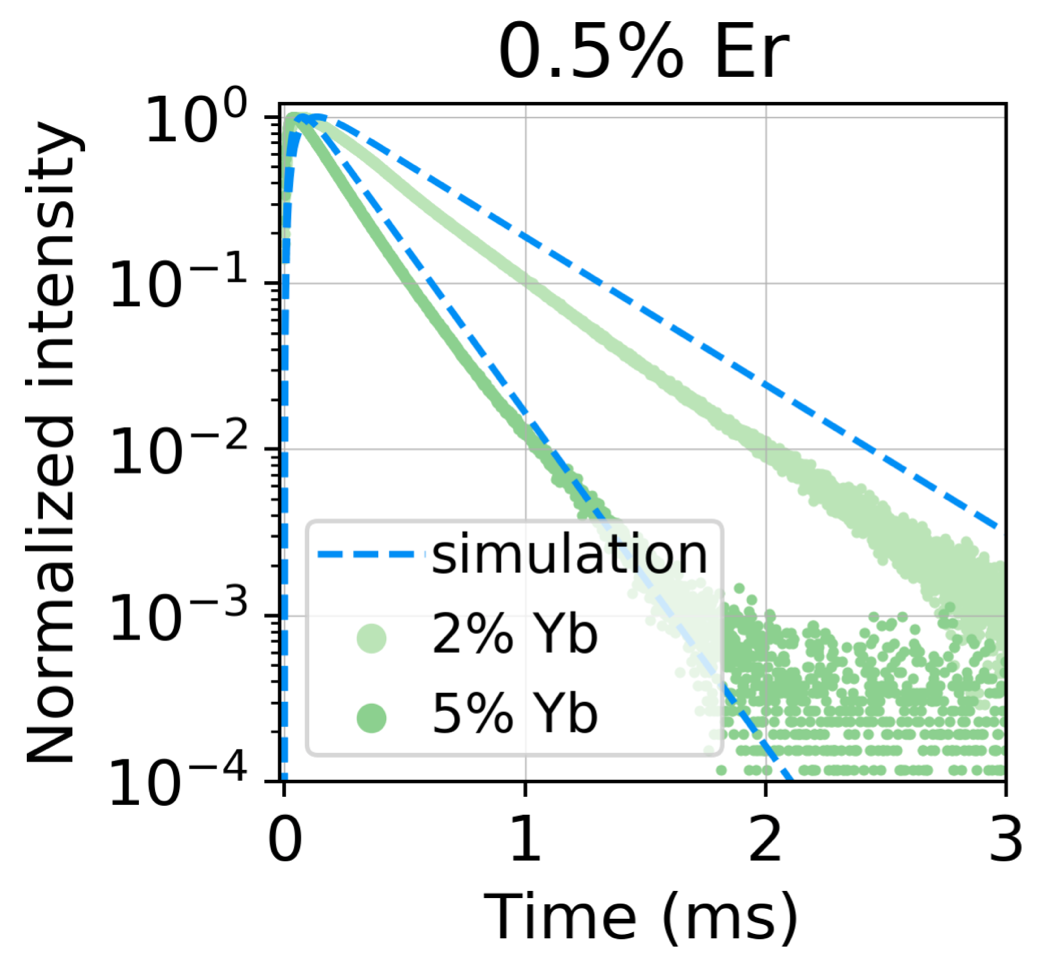}
	\caption{Green-emitting levels decays of co-doped nanoparticles following excitation at 980 nm and the associated modeling.}
	\label{fig:si_sim_uc}
\end{figure}

\subsection{Variation of excited state lifetimes under NIR excitation as a function of Er concentration}

Figure~\ref{fig:si_sim_yb} presents the relaxation rate $K_{1'}$ as a function of Er concentration, for nanoparticles co-doped with 2\% Yb and 5\% Yb. The modeling shows that increasing Er concentration leads to shortening of \Ybt\ lifetime, confirming that \Ert\ ions act as quenching centers for \Ybt\ ions.

\begin{figure}[h]\centering
	\includegraphics[width=0.6\textwidth]{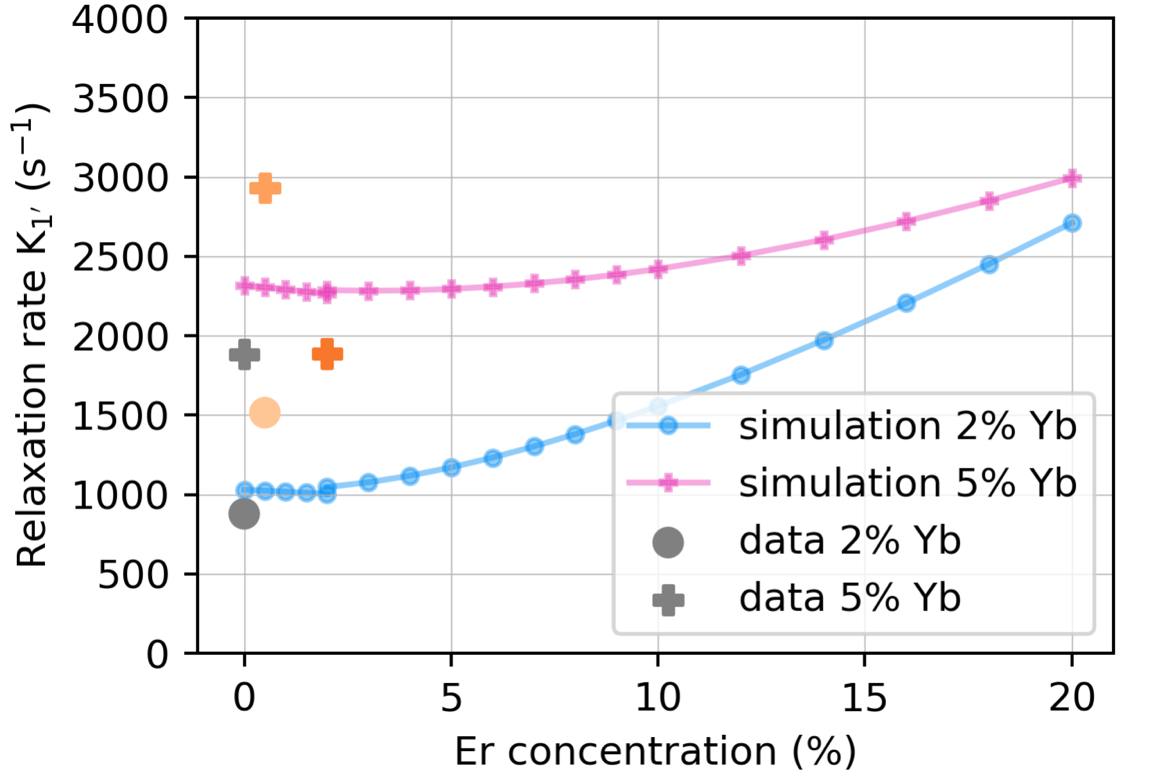}
	\caption{Relaxation rate of the \Ybt\ excited sate level following excitation at 980 nm for co-doped nanoparticles with 2\% Yb and 5\% Yb as a function of Er concentration.}
	\label{fig:si_sim_yb}
\end{figure}

Figure~\ref{fig:si_sim_green} and Figure~\ref{fig:si_sim_red} present the relaxation rate $K_{5}$ of the green-emitting levels ($^2$H$_{11/2}$,$^4$S$_{3/2}$) and the relaxation rate $K_4$ of the red-emitting level ($^4$F$_{9/2}$), as a function of Er concentration, for nanoparticles co-doped with 2\% Yb and 5\% Yb. The modeling shows that relaxation rates of both levels increases with Er increasing.

\begin{figure}[h]\centering
	\includegraphics[width=0.6\textwidth]{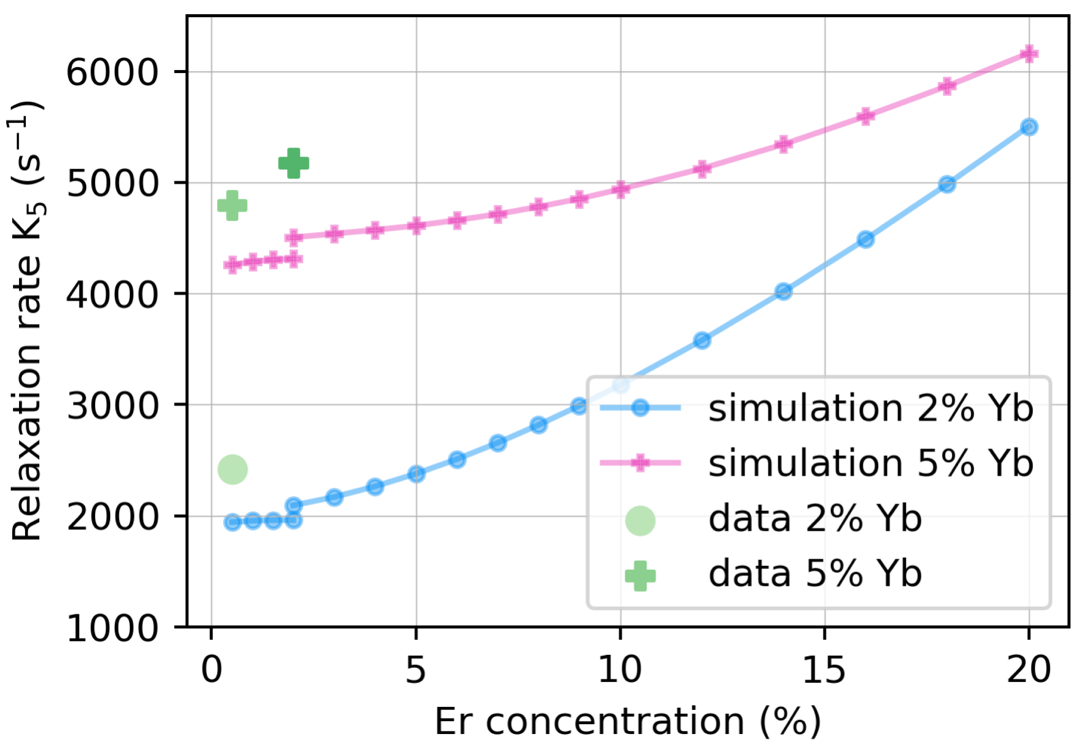}
	\caption{Relaxation rate of the $^2$H$_{11/2}$,$^4$S$_{3/2}$ levels following excitation at 980 nm for co-doped nanoparticles with 2\% Yb and 5\% Yb as a function of Er concentration.}
	\label{fig:si_sim_green}
\end{figure}

\begin{figure}[h]\centering
	\includegraphics[width=0.6\textwidth]{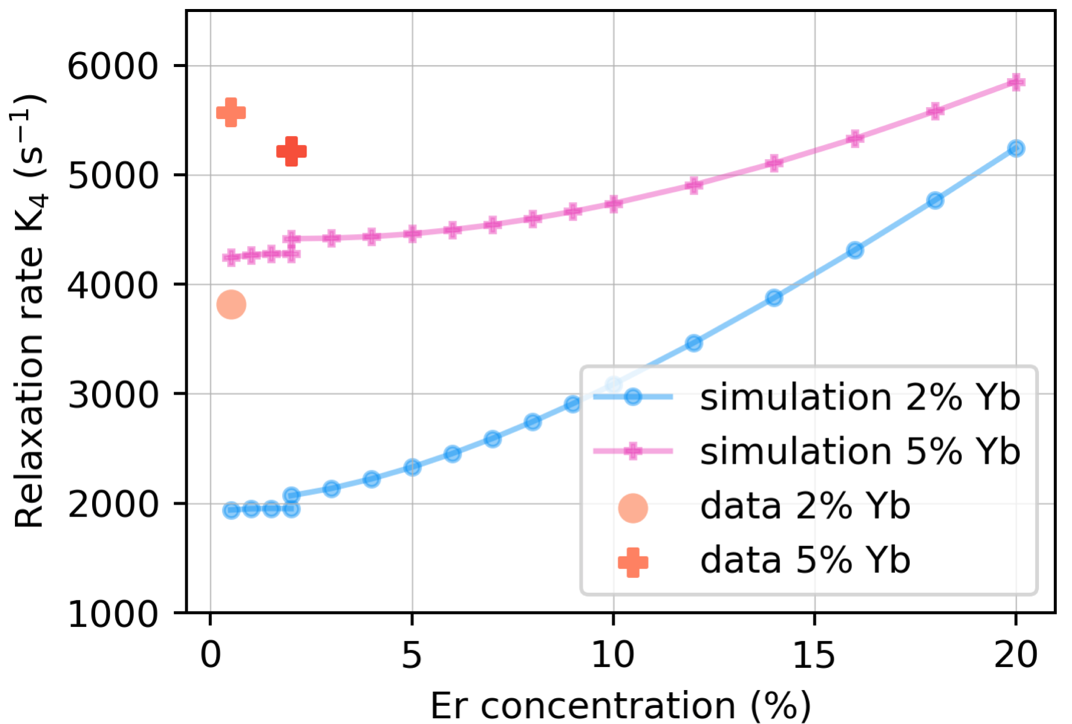}
	\caption{Relaxation rate of the $^4$F$_{9/2}$ level following excitation at 980 nm for co-doped nanoparticles with 2\% Yb and 5\% Yb as a function of Er concentration.}
	\label{fig:si_sim_red}
\end{figure}

\end{suppinfo}

\clearpage
\bibliography{bibliography}

\end{document}